\documentclass[10pt, conference, compsocconf]{IEEEtran}
\IEEEoverridecommandlockouts
\ifCLASSINFOpdf
\else
\fi

\usepackage{caption2}

\usepackage{graphicx}
\usepackage{subfigure}
\usepackage{amsmath,amsthm,amssymb,bm}
\usepackage{algorithm} 
\usepackage{algorithmic} 
\usepackage{multirow}
\usepackage{epstopdf}
\usepackage{setspace}
\usepackage{balance}

\theoremstyle{proposition}
\newtheorem{proposition}{Proposition}
\theoremstyle{proposition}
\newtheorem{property}{Property}
\theoremstyle{definition}
\newtheorem{definition}{Definition}

\begin{document}
%
\title{Privacy Preserving Social Network Publication Against Mutual Friend Attacks}

\author{\IEEEauthorblockN{Chongjing Sun\IEEEauthorrefmark{2},
Philip S. Yu\IEEEauthorrefmark{3},
Xiangnan Kong\IEEEauthorrefmark{3} and
Yan Fu\IEEEauthorrefmark{2}}
\IEEEauthorblockA{\IEEEauthorrefmark{2}Web Science Center, School of Computer Science and Engineering\\
University of Electronic Science and Technology of China,
Chengdu, China, 611731 }
\IEEEauthorblockA{\IEEEauthorrefmark{3}Department of Computer Science, University of Illinois at Chicago, Chicago, IL 60612\\
}
\IEEEauthorblockA{Email:chingsun00@gmail.com, psyu@cs.uic.edu, xkong4@cs.uic.edu, fuyan@uestc.edu.cn}
\thanks{Yan Fu is the corresponding author.}
\thanks{This research work was supported in part by the National Natural Science Foundation of China under Grant No.61003231 and No.60973120, the research funds for central universities under grant No. ZYGX2012J085, and US NSF through grants CNS-1115234,  DBI-0960443, and OISE-1129076.}
}


\maketitle

\begin{abstract}
Publishing social network data for research purposes has raised serious concerns for individual privacy. There exist many privacy-preserving works that can deal with different attack models. In this paper, we introduce a novel privacy attack model and refer it as a mutual friend attack. In this model, the adversary can re-identify a pair of friends by using their number of mutual friends.
To address this issue, we propose a new anonymity concept, called \emph{k}-NMF anonymity, i.e., \emph{k}-anonymity on the number of mutual friends, which ensures that there exist at least \emph{k}-1 other friend pairs in the graph that share the same number of mutual friends.
We devise algorithms to achieve the \emph{k}-NMF anonymity
while preserving the original vertex set in the sense that we allow the occasional addition but no deletion of vertices.
Further we give an algorithm to ensure the \emph{k}-degree anonymity in addition to the \emph{k}-NMF anonymity.
The experimental results on real-word datasets demonstrate that our approach can preserve the privacy and utility of social networks effectively
 against mutual friend attacks.
\end{abstract}


%
\IEEEpeerreviewmaketitle

\section{Introduction}
With the advance on mobile and Internet technology, more and more information is recorded by social network applications, such as Facebook and Twitter.
The relationship information in social networks attracts researchers from different academic fields.
As a consequence, more and more social network datasets were published for research purposes \cite{Stanford_Large_Network}.
The published social network datasets may incur the privacy invasion of some individuals or groups. With the increasing concerns on the privacy, many works have been proposed for the privacy-preserving social network publication \cite{B.Zhou:privacy_survey, X.Wu:privacy_survey}.

\begin{figure}[!t]
\centering
\includegraphics[width=0.45\textwidth]{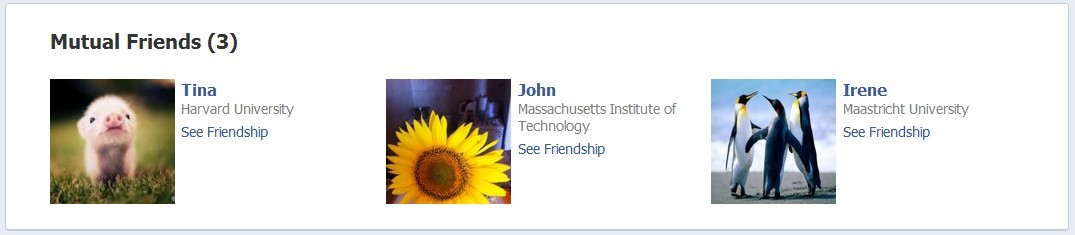}
\includegraphics[width=0.45\textwidth]{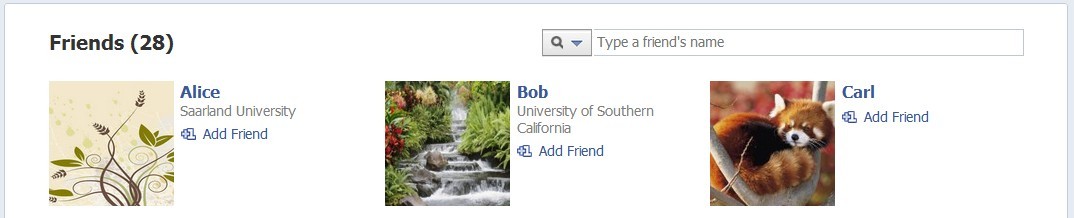}
\caption{Friend lists on Facebook}
\label{facebook}
\end{figure}

\begin{figure}[!t]
\centering
\subfigure[]{
\label{mfa}
\begin{minipage}[t]{0.13\textwidth}
      \centering
      \includegraphics[height=0.7in]{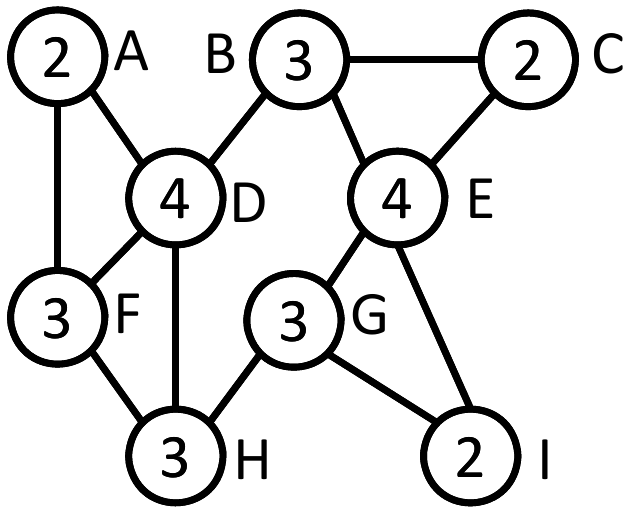}
\end{minipage}}
\subfigure[]{
\label{osn}
\begin{minipage}[t]{0.13\textwidth}
      \centering
      \includegraphics[height=0.7in]{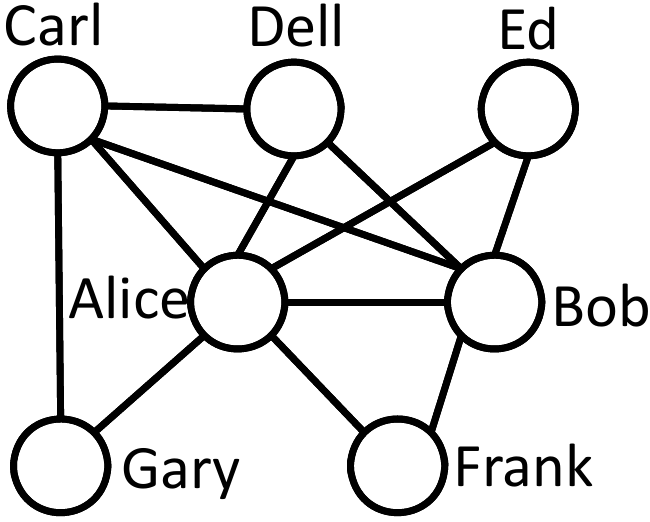}
\end{minipage}}
\subfigure[]{
\label{nan}
\begin{minipage}[t]{0.13\textwidth}
      \centering
      \includegraphics[height=0.7in]{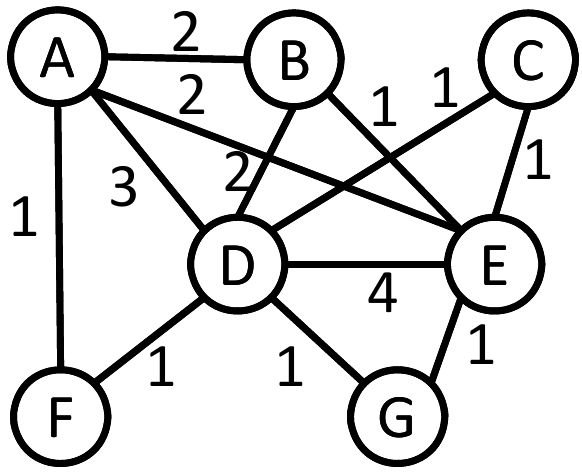}
\end{minipage}}
\caption{Mutual friend attack in a social network}
\label{example_mutual}
\end{figure}

Tai and Yu proposed the friendship attack model \cite{C.Tai:friendship_attack}, which addressed the issue that
an attacker can find out not only the degree of a person, but also the degree of his friend.
It solves the attacks based on the degrees of two connected vertices.
But it is not sufficient to just protect against the friendship attack as there are more information available on the social network.
For example, the graph in Fig. \ref{mfa} is a $k^2$-degree anonymized graph with $k=2$.
If an attacker can obtain the number of mutual friends between two connected vertices, he still can identify $(D,F)$ from other friend pairs, as only $(D,F)$ has 2 mutual friends. This will be explained in more details later.
In most social networking sites, such as Facebook, Twitter, and LinkedIn, the adversary can easily get the number of mutual friends of two individuals linked by a relationship. As shown in Figure \ref{facebook}, one can directly see mutual friend list shared with one of his friends on Facebook.
Usually, the adversary can get the friend lists of two individuals from Facebook, such as the friend list in Figure \ref{facebook},
and then get the number of mutual friends by intersecting their friend lists.

In this paper, we introduce
a new relationship attack model based on the number of mutual friends of two connected individuals, and refer it as a \emph{mutual friend attack}.
Figure \ref{example_mutual} shows an example of the mutual friend attack.
The original social network
\emph{G} with vertex identities is shown in
Figure \ref{osn},
and can be naively anonymized as the network \emph{G'} shown in Figure \ref{nan} by removing all
individuals' names.
The \emph{number} on each edge in \emph{G'}
represents
the number of mutual friends of the two end vertices.
Alice and Bob are friends, and their mutual friends are Carl, Dell, Ed and Frank.
So the number of mutual friends of Alice and Bob is 4. After obtaining this information, the adversary can uniquely re-identify the edge $(D,E)$ is $(Alice, Bob)$. Also, $(Alice, Carl)$ can be uniquely re-identified in \emph{G'}. By combining $(Alice, Bob)$ and $(Alice, Carl)$, the adversary can uniquely re-identify individuals Alice, Bob and Carl. This simple example illustrates that it is possible for the adversary to re-identify an edge between two individuals and maybe indeed identify the individuals when he can get the number of mutual friends of individuals. Note that we do not consider the mutual friend number of two nodes if they are not connected. For convenience, we say the \emph{number of mutual friends of two nodes} connected by an edge $e$ as the \emph{number of mutual friends of $e$}.

In order to protect the privacy of relationship from the mutual friend attack, we introduce a new privacy-preserving model, \emph{k}-anonymity on the number of mutual friends (\emph{k}-NMF Anonymity). For each edge \emph{e}, there will be at least \emph{k}-1 other edges with the same number of mutual friends as \emph{e}. It can be guaranteed that the probability of an edge being identified is not greater than 1/\emph{k}. We propose algorithms to achieve the \emph{k}-NMF anonymity for the original graph
 while preserving the original vertex set in the sense that we allow the occasional addition but no deletion of vertices.
 By preserving the original vertex set, various analysis on the anonymized graph, such as identifying vertices providing specific roles like centrality vertex, influential
 vertex, gateway vertex, outlier vertex, etc., will be more meaningful.
 The experimental results on real datasets show that our approaches can preserve much of the utility of social networks against mutual friend attacks.

\noindent\textbf{Related Work}.
Backstorm et al. \cite{Backstrom:Wherefore_art_thou} pointed out that simply removing identities of vertices cannot guarantee privacy. Many works have been done to prevent the vertex re-identification with the vertex degree.
Liu et al. \cite{K.Liu:k-anonymization} studied the \emph{k}-degree anonymization which ensures that for any node \emph{v} there exist at least \emph{k}-1 other vertices in the published graph with the same degree as \emph{v}.
Tai et al. \cite{C.Tai:friendship_attack} introduced a friendship attack, in which the adversary uses the degrees of two end vertices of an edge to re-identify victims.
Associated with community identity for each vertex, in \cite{C.Tai:structural_diversity} they proposed the $k$-structural diversity anonymization, which guarantees the existence of at least \emph{k} communities containing vertices with the same degree for each vertex. As these works only focus on the vertex degree, they cannot achieve the \emph{k}-NMF anonymity, which focuses on the number of common neighbors of two vertices.

\begin{figure}[!t]
\centering
\subfigure[]{
\label{2_automorp}
\begin{minipage}[t]{0.2\textwidth}
      \centering
      \includegraphics[height=0.6in]{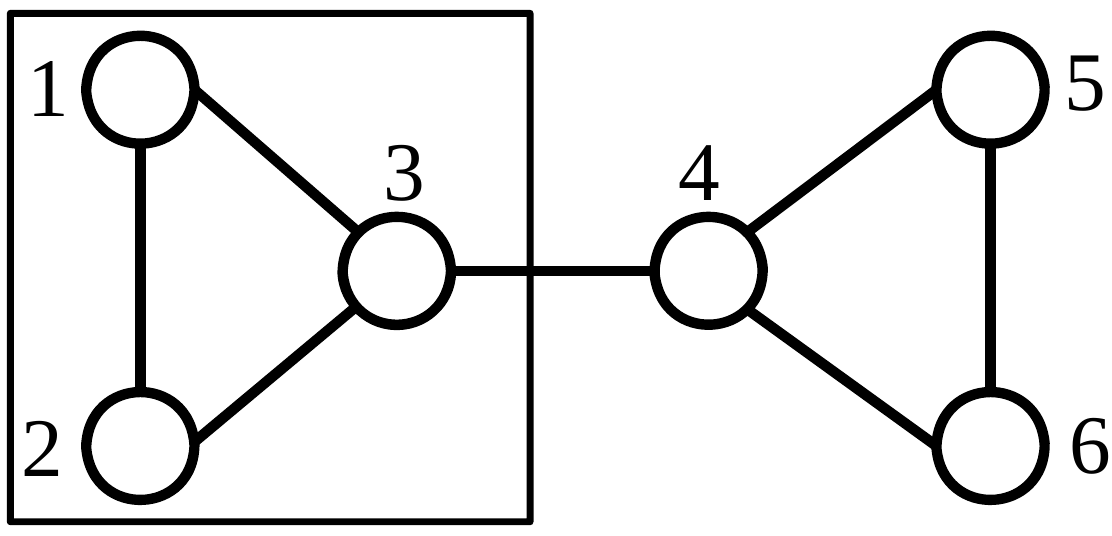}
\end{minipage}}
\subfigure[]{
\label{different_edges_addition}
\begin{minipage}[t]{0.12\textwidth}
      \centering
      \includegraphics[height=0.6in]{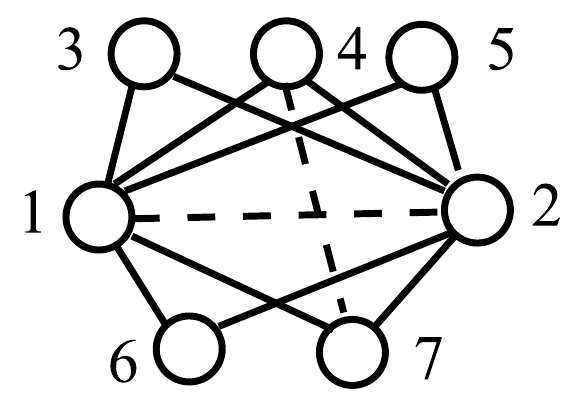}
\end{minipage}}
\caption{Examples of the k-NMF anonymization}
\label{Fig1}
\end{figure}

Many works have also been done to prevent the vertex re-identification based on the subgraph structural information.
Zhou and Pei \cite{Zhou:neighborhood_attacks} proposed a solution to battle the adversary's 1-neighborhood attacks.
Cheng et al. \cite{James:k_iso} proposed the \emph{k}-isomorphism model, which disconnects the original graph into $k$-isomorphic subgraph.
To protect against multiple structural attacks, Zou et al. \cite{Zou:K-Automorphism} proposed the \emph{k}-automorphism model, which converts the original network into a $k$-automorphic network.
But it does not prevent the mutual friend attack. The network in Figure \ref{2_automorp} satisfies the 2-automorphism, but the edge $(3,4)$ is not protected under the mutual friend attack.
This is because the edge $(3,4)$ does not have mutual friends while all the others have one.
Wu et al. \cite{Wu:k-Symmetry} proposed the $k$-symmetry model, which gets a $k$-automorphic network by orbit copying. All these algorithms need to introduce many new vertices and adjust many edges to achieve their targets. Therefore, the utility of the original graph will be decreased too much.
In any case, these works are aimed at different types of attack model from ours as illustrated in Figure \ref{2_automorp}.

Hay et al. \cite{Hay:Structural} proposed a generalizing method for anonymizing a graph, which partitions the vertices and summarizes the graph at the partition level. Other works focus on the problem of link disclosure, which decides whether there exists a link between two individuals. It is different from the relationship re-identification introduced in this section.

\noindent\textbf{Challenges}.
As the $k$-NMF anonymity model is  more complicated than the $k$-degree anonymity model, more challenges need to be handled.
First, adding or removing a different edge may affect a different number of edges on their mutual friends.
In the $k$-degree anonymity model, the adversary attacks using the degree of the vertex.
Adding an edge only increase the degrees of the two end vertices of this edge.
In the $k$-NMF anonymity model,
the adversary attacks using the number of mutual friends.
Adding an edge can increase the
numbers of mutual friends of many edges.
In Figure \ref{osn}, adding an edge between Dell and Frank will affect the NMFs of
(Dell, Alice), (Frank, Alice), (Dell, Bob), (Frank, Bob), and (Dell, Frank).
Second, we need to provide a criterion on choosing where to add or delete the edge while considering the utility of the graph. Since we aim to preserve the vertex set,
we cannot add a vertex to connect an edge.
In fact, we map the $k$-NMF anonymization
problem into an edge anonymization problem
in contrast to the vertex anonymization problem in the $k$-degree anonymization. Edges are anonymized one by one.
Adding or deleting an edge should not destroy the anonymization of the already anonymized edges.
To anonymize an edge, we can get many candidate edge operations and need to choose the best one.
Besides, we need to consider the impact of the newly added edges on the number of mutual friends.

\noindent\textbf{Contributions}.
Our contributions can be summarized as follows. (1) We introduce the $k$-NMF problem and formulate it as an edge weight anonymization problem where the edge weight is the NMF of the two end vertices. (2) We explore the geometry property of the graph to devise effective anonymization algorithms while preserving the vertex set to achieve better utility.
(3) For the edge addition, we use the breadth-first manner to preserve utility. We also introduce the maximum mutual friend criterion to break the tie on selecting candidate vertex to connect. (4) For the edge deletion, we explore the triangle linking property to delete edges between vertices already belonging to a triangle connection in the network to avoid repeated re-anonymization of edges. (5) We devise an algorithm which can anonymize the \emph{k}-NMF anonymized graph to simultaneously satisfy the \emph{k}-degree anonymity, while preserving the vertex set. (6)
The empirical results on real datasets show that our algorithms perform well in anonymizing the real social networks.

The rest of the paper is organized as follows.
We define the problem and design algorithms to solve it in section 2 and 3.
We conduct the experiments on real data sets and conclude in Section 4 and 5.

\section{Problem definition}\label{sec: probelm definition}
In this paper, we model a social network as an undirected simple graph $G(V, E)$, where \emph{V} is a set of vertices representing the individuals, and $E\subseteq V\times V$ is the set of edges representing the relationship of individuals.
\vspace{-0.1cm}
\begin{definition}\label{def:mfne}\textbf{\emph{The NMF of an edge.}}
For an edge $e$ between two vertices $v_1$ and $v_2$ in a graph $G(V,E)$, i.e., $v_1, v_2\in V$, $e\in E$ and $e=(v_1, v_2)$, the number of mutual friends of the edge $e$ is the number of mutual friends of $v_1$ and $v_2$.
\end{definition}

\begin{figure}[!t]
\centering
\subfigure[3-NMF]{\label{f2-1}
\includegraphics[width=0.11\textwidth]{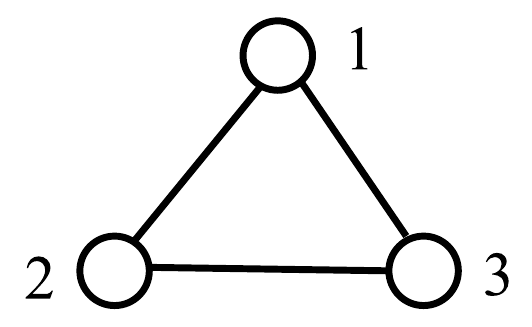}}
\subfigure[6-NMF]{\label{f2-2}
\includegraphics[width=0.12\textwidth]{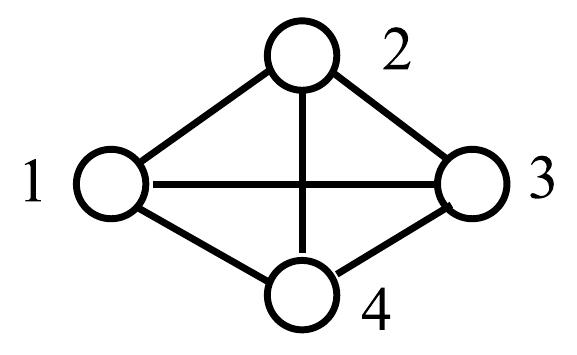}}
\subfigure[4-NMF]{\label{f2-3}
\includegraphics[width=0.12\textwidth]{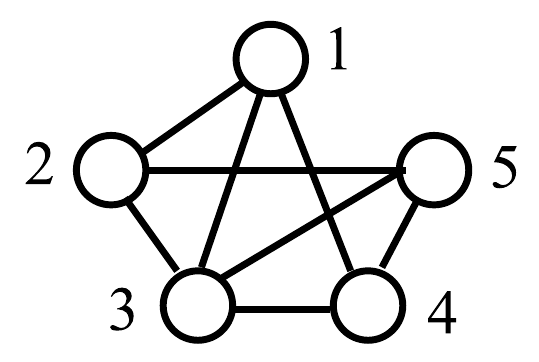}}
\caption{Examples of \emph{k}-NMF anonymous graph}
\vspace{-0.4cm}
\label{Fig2}
\end{figure}

\vspace{-0.3cm}
Let $\bm{f}$ be the \emph{number sequence of mutual friends} for $G$, in which entries are sorted in descending order, i.e., $\bm{f}_1\geq \bm{f}_2\geq ...\geq \bm{f}_m$.
Let $\bm{l}$ be the list of edges corresponding to $\bm{f}$, i.e., $\bm{f}_i$ is the NMF of the edge $\bm{l}_i$.
For example, in Figure \ref{f2-3}, $\bm{f}=\{2,2,2,2,1,1,$ $1,1\}$,
and $\bm{l}=\{(v_1,v_3),(v_2,v_3),(v_3,v_4), (v_3,v_5),(v_3,v_4), (v_3,v_5), (v_1,v_2),$ $(v_1,v_4),(v_2, v_5), (v_4,v_5)\}$.
Similar to the power law distribution of the vertex degree \cite{Faloutsos:power_law_degrees}, the NMF also has the same property \cite{Vinko:edge_multiplicities}.

\vspace{-0.1cm}
\begin{property}\label{pro:power}\textbf{\emph{Scale free distribution of NMFs}} \emph{\cite{Vinko:edge_multiplicities}}\textbf{.}
The NMFs of edges in the large social network often have a scale-free distribution, which means that the
distribution follows a power law or at least asymptotically.
\end{property}

\vspace{-0.2cm}
\begin{definition}\label{def:mfa}\textbf{\emph{Mutual friend attack}}.
Given a social network $G(V, E)$ and the anonymized network $G'(V',E')$ for publishing. For an edge $e\in E$, the adversary can get the number $\bm{f}_e$ of mutual friends of $e$. Mutual Friend Attack will identify all \emph{candidate edges} $e'\in E'$ with the number $\bm{f}_{e'}$ of mutual friends as $\bm{f}_e$.
\end{definition}

\vspace{-0.3cm}
Suppose that the candidate edge set of an edge $e$ is $E'_{e}=\{e'|e'\in E', \bm{f}_{e'}=\bm{f}_e\}$. An adversary re-identifies the edge $e$ with high confidence if the number of candidate edges is too small. Hence, we set a threshold $k$ to make sure that for each edge $e\in E$, the number of candidate edges is no less than $k$, i.e., $|E'_{\bm{f}_e}|\geq k$. We define the \emph{k}-anonymous sequence before defining the \emph{k}-NMF anonymous graph.

\vspace{-0.1cm}
\begin{definition}\label{def:kseq}\textbf{\emph{\textbf{k}-anonymous sequence}}\cite{K.Liu:k-anonymization}.
A sequence vector $\bm{f}$ is $k$-anonymous, if for any entry with value as $v$, there exist at least $k-1$ other entries with value as $v$.
\end{definition}

\vspace{-0.2cm}
\begin{definition}\label{def:kmfn}\textbf{\emph{k}-NMF}.
A graph $G'(V',E')$ is $k$-NMF anonymous if the number sequence $\bm{f}'$ of mutual friends of edges in $G'$ is a $k$-anonymous sequence.
\end{definition}

\vspace{-0.3cm}
Definition \ref{def:kmfn} states that for each edge $e\in E$, the number of candidate edges in $G'$ is no less than $k$. Consider the graphs in Figure \ref{Fig2} as an example. There are three edges in Figure \ref{f2-1}, and the NMFs of all these edges are equal to 1. Hence, this graph is a 3-NMF anonymous graph. As the six edges in the graph of Figure \ref{f2-2} have 2 mutual friends, this graph is a 6-NMF anonymous graph. The graph in Figure \ref{f2-3} has four edges $(v_1,v_3),(v_2,v_3),(v_3,v_4),(v_3,v_5)$ with the NMF as 2, and the NMFs of other four edges are equal to 1. Hence, this graph is a 4-NMF anonymous graph. Some properties on the number of mutual friends in the graph are described as follows.

\vspace{-0.1cm}
\begin{proposition}\label{prop3}
Given a graph $G(V,E)$, the number of mutual friends of an edge $e\in E$ is equal to the number of triangles containing $e$ in $G$ .
\end{proposition}

\vspace{-0.3cm}
Take the graph in Figure \ref{f2-3} as an example. The mutual friends of vertices $v_2$ and $v_3$ are $v_1$ and $v_5$, so the number of mutual friends of the edge $e=(v_2, v_3)$ is 2. It is equal to the number of triangles containing $e$.  These triangles are $(v_1,v_2,v_3)$ and $(v_2,v_3,v_5)$.

\vspace{-0.1cm}
\begin{proposition}\label{prop4}
Let $G(V,E)$ be a graph and $f$ be the number sequence of mutual friends of edges in $G$, where $|E|=m$. Then $\sum_{i=1}^m\bm{f}_i=3n_{\vartriangle}$, where $n_{\vartriangle}$ is the number of triangles in $G$ and $\bm{f}_i$ is the number of mutual friends of the $i$-th edge.
\end{proposition}

\vspace{-0.3cm}
Different from the degree sequence in previous work \cite{K.Liu:k-anonymization}, which can maintain the number of entries in the sequence, the number sequence of mutual friends will have more entries added into it when new edges are added into the graph. Besides, according to Propositions \ref{prop3} and \ref{prop4}, the number of mutual friends is related to the number of triangles in the graph. Therefore, adding one edge will affect the NMF of many edges, and adding a different edge may affect the NMF of a different number of edges. This can be illustrated by an example shown in Figure \ref{different_edges_addition}. After we add the edge $(1,2)$, the NMFs of all ten edges increase by one. If we add the edge $(4,7)$, only the NMFs of edges $(1,4),(1,7),(2,4)$, and $(2,7)$ increase by one. Therefore, one cannot anonymize a graph by simply minimizing the number of changed edges.

\noindent\textbf{Anonymized Triangle Preservation Principle (ATPP)}. In our algorithms, we anonymize the edges in the graph one by one. An \emph{anonymized triangle} is a triangle with some edges already anonymized in the process of the graph anonymizing.
The \emph{Anonymized Triangle Preservation Principle} aims to preserve the anonymized triangles containing already anonymized edges. It means that we neither create some additional anonymized triangles via edge addition nor destroy any via edge deletion.

Creating (destroying) a triangle containing an already anonymized edge by edge addition (deletion) will increase (decrease) the NMF of this edge, indeed destroy the anonymization of this edge. This leads to repeatedly anonymization of this edge.
By preserving the anonymized triangles, we can avoid this problem during the anonymization process.

\vspace{-0.15cm}
\begin{definition}\label{def:problem}\textbf{\emph{k}-NMF anonymization problem}.
Given a graph $G(V,E)$ and an integer $k$, the problem is to anonymize the graph $G$ to a $k$-NMF anonymous graph $G'$ with edge addition and deletion, such that the vertex set of the original graph $G$ is preserved.
\end{definition}

\vspace{-0.25cm}
\section{\emph{k}-NMF anonymization approach}
In the above section, we found that changing one edge may affect the NMFs of other edges. To handle this challenge, we utilize the scala free distribution property shown in Property \ref{pro:power}, and introduce the principle of preserving the anonymized triangles. By exploring the geometry property of the graph, we devise two effective anonymization algorithms to preserve the utility while satisfying the $k$-NMF anonymity.

\vspace{-0.2cm}
\subsection{Algorithm ADD}\label{sub:add}
In this subsection, we aim to anonymize the original graph only by edge addition.
We organize edges into groups, and anonymize the edges in the same group to have the same NMF.
The $k$-anonymity requires there exist at least $k$ edges in a group. Property \ref{pro:power} states that the NMFs of edges in large social networks follow a scala free distribution. Hence, only a small number of edges have a high NMF. We first anonymize these edges, and many edges with low NMF do not need to be processed.

Suppose the original graph is $G(V,E)$ and the gradually anonymized graph is $G'(V',E')$. Initially, we sort the NMF sequence $\bm{f}$ in descending order and construct the corresponding edge list $\bm{l}$ as described in Section \ref{sec: probelm definition}. We mark all edges as ``unanonymized", and then anonymize the edges one by one. Iteratively, we start a new group $GP$ with the group NMF, $g_f$, equal to the NMF of the first unanonymized edge in $\bm{l}$.
Then we select the edges with NMF equal to $g_f$ and mark them as ``anonymized".
We iteratively select the first unanonymized edge in $\bm{l}$
and anonymize it by adding edges to increase its NMF to $g_f$. After anonymizing this edge, we mark it as ``anonymized" and put it into $GP$.
Adding new edges affects the NMF of some other edges, and these new edges will be added into $\bm{f}$ and $\bm{l}$. Hence we resort the sequences $\bm{f}$ and $\bm{l}$ after each edge is anonymized. Algorithm 1 shows the detailed description of the ADD algorithm. Next, we consider when we start another new group.

\subsubsection{Group edges}
An intuitive method, named \textbf{IntuitGroup}, starts another group when the number of edges in the group $GP$ is equal to $k$. Alternatively, to consider the anonymization cost,
we propose a greedy method to decide when we start another group after $|GP|\geq k$, named \textbf{GreedyGroup}.
Suppose that $\bm{f}^{(u)}\subseteq \bm{f}$ is the NMF sequence corresponding to the unanonymized edge list $\bm{l}^{(u)}\subseteq \bm{l}$. Notice that $\bm{f}^{(u)}$ and $\bm{l}^{(u)}$ are dynamically updated with $\bm{f}$ and $\bm{l}$ after anonymizing each edge.
Similar to the consideration in \cite{K.Liu:k-anonymization},
after putting $k$ edges into $GP$,
GreedyGroup iteratively checks whether it should merge the edge $\bm{l}^{(u)}_{1}$ into $GP$ or start another group.
The decision is made according
to the following two costs based on the number of added mutual friends in Eq.(\ref{formlar1}) and Eq.(\ref{formlar2}).
\begin{equation}\label{formlar1}
C_{merge}=(g_f-\bm{f}^{(u)}_{1})+I(\bm{f}^{(u)}_{2},\bm{f}^{(u)}_{k+1})
\end{equation}
\vspace{-0.4cm}
\begin{equation}\label{formlar2}
C_{new} = I(\bm{f}^{(u)}_{1},\bm{f}^{(u)}_{k})
\end{equation}
where $I(\bm{f}^{(u)}_{i},\bm{f}^{(u)}_{j})=\sum_{l=i}^{j}(\bm{f}^{(u)}_{i}-\bm{f}^{(u)}_{l})$.

For Eq.(\ref{formlar1}), we put $\bm{l}^{(u)}_{1}$ into $GP$. $\bm{l}^{(u)}_{1}$ has $\bm{f}^{(u)}_{1}$ mutual friends,
so we need to add $g_f-\bm{f}^{(u)}_{1}$ mutual friends for anonymizing $\bm{l}^{(u)}_{1}$. To satisfy $k$-anonymity, we need to put at least $k$ edges into a new group $GP'$. Hence we put edges
$\bm{l}^{(u)}_{2},...,\bm{l}^{(u)}_{k+1}$ into $GP'$.
As we only adding edges, the group NMF of $GP'$ is the maximum NMF among $\bm{f}^{(u)}_{2},...,\bm{f}^{(u)}_{k+1}$, i.e., $\bm{f}^{(u)}_{2}$.
To anonymize $\bm{l}^{(u)}_{i}$, $\bm{f}^{(u)}_{2}-\bm{f}^{(u)}_{i}$ mutual friends need to be added. For Eq.(\ref{formlar2}), we put $\bm{l}^{(u)}_{1},...,\bm{l}^{(u)}_{k}$ into a new group $GP'$,
and the group NMF of $GP$ is $\bm{f}^{(u)}_{1}$.
Hence $C_{merge}$ is the cost for anonymizing $k+1$ edges while $C_{new}$ is for $k$ edges.
So if $C_{merge}$ is less than $C_{new}$, we anonymize $\bm{l}^{(u)}_1$ and merge it into $GP$, and check the next unanonymized edge.
Otherwise we start another new group with $\bm{l}^{(u)}_1$.

For each edge $e$, GreedyGroup looks ahead at $O(k)$ other edges to decide whether merging $e$ with this group or starting a new group. Therefore, the time complexity of GreedyGroup is $O(k|E|)$.

\begin{figure}[!t]
\centering
\includegraphics[width=0.48\textwidth]{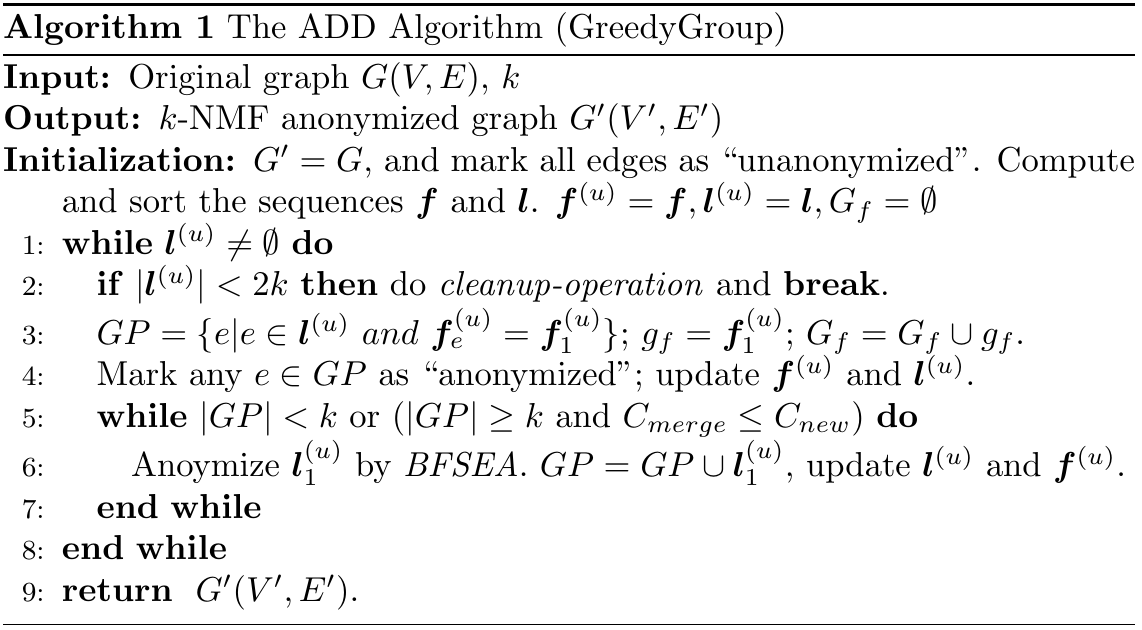}
\label{alg:ADD}
\vspace{-0.5cm}
\end{figure}

\subsubsection{Cleanup-operation}
\vspace{-0.1cm}
In each iteration of the ADD algorithm,
it checks the number of unanonymized edges, $n_{u}$. If $n_{u}<2k$, the ramaining edges are put into a group; and if $n_{u}<k$, $k-n_{u}$ edges needed to be added following the ATPP, so these $k$ edges can form a group. New vertices will be added into the graph if the ATPP cannot be satisfied.

Next, we anonymize the edges $E_u$ in this group.
Usually, we set the group NMF as the largest NMF among unanonymized edges, denoted as $g_f$.
Then we sum the difference as $sd=\sum_{e\in E_u}(g_f-\bm{f}_e)$, where $\bm{f}_e$ is the number of mutual friends of the edge $e$.
If $sd>=k/2$, then we add $sd$ nodes and $2\cdot sd$ edges into the graph. That means that for each unanonymized edge $e$, we add $g_f-\bm{f}_e$ vertices and link them with the two end vertices of $e$.
As all the newly added $(2\cdot sd>=k)$ edges have only one mutual friend, they can form a new group. Then we mark the new edges as ``anonymized" and achieve the task. If $sd<k/2$, then we enlarge the group NMF $g_f=g_f+1$, and repeat the above process. By the \emph{clean-up operation}, we can successfully anonymize the original network at the last step of the anonymization process.

\vspace{-0.2cm}
\subsection{BFS-based Edge Anonymization(BFSEA)}
In this section, we consider how to anonymize an edge by edge addition while preserving the utility. There are three challenges to increase the NMF of an edge via adding edges. First, the added edge should not affect the NMF of already anonymized edges.
Secondly, the added edge should minimize the effect on the utility of the graph.
Thirdly, the NMF of the newly added edges should not disrupt the current anonymization process which is progressing in descending order of the NMF value.

Before anonymizing an edge $(u,v)$, the $ADD$ algorithm has created some anonymized groups and got a set $G_f$ containing the group NMFs of these groups. Let $g_f$ be the NMF of the current group $GP$, and we put $g_f$ into $G_f$.
Anonymizing the edge $(u,v)$ means that we need to increase the NMF of $(u,v)$ to the current group NMF $g_f$, i.e. we need to create some new triangles containing this edge. Then we try to find some candidate vertices and add new edges to create new triangles. Considering the utility of the graph, we find the candidate vertices based on the Breadth First Search (BFS).

From the nodes $u$ and $v$, \emph{BFS-based Edge Anonymization} traverses the graph in a breadth-first manner. For the $i$-hop neighbors of $u$ and $v$, represented by $neig_i(u)$ and $neig_i(v)$, \emph{edge anonymization} finds the candidate vertices from $neig_i(u)\cup neig_i(v)$ and iteratively link the best one with $u$ or $v$ to create a new triangle.
We formulize the NMF of the edge $(u,v)$ as $nmf(u,v)$.

\subsubsection{Candidates generation}
We search the candidate vertices for edge $(u,v)$ in a BFS manner. In the $i$-hop neighbors of $u$ and $v$, many vertices cannot be the candidate vertices as violating the ATPP. The vertices $w$ need to satisfy the following conditions to be the candidates in the set $CV_i$.

\renewcommand{\theenumi}{\alph{enumi}}
\begin{enumerate}\vspace{-0.2cm}
  \item $w\in neig_i(u)\cup neig_i(v)$.
  \item $(w,u,v)\neq\triangle$.
  \item $\forall x\in\{u,v\}~and~z\in V'$, if $(w,x)\not\in E',(w,z)\in E'~and~(x,z)\in E'$, then $(x,z)$ and $(w,z)$ must be unanonymized. \vspace{-0.2cm}
\end{enumerate}

Condition b) states that $(u,v,w)$ is not a complete triangle, which needs to add edges to create a new triangle. This mainly focus on the case when $i=1$, where $w$ may links with both $u$ and $v$. Condition c) follows the ATPP, which guarantees that there will be no effect on the already anonymized edges.

\subsubsection{Candidates selection}
After getting all the candidate vertices satisfying the conditions, we can add new edges between $u$,$v$ and $w\in CV_i$ to increase the NMF of $(u,v)$. We iteratively select a vertex from $CV_i$ to increase the NMF of $(u,v)$ until $nmf(u,v)$ reaches $g_f$ or $CV_i$ is empty. If $nmf(u,v)=g_f$,
this edge is anonymized successfully.

In each iteration, we need to select the best one which can preserve the most utility of the graph.
Based on the link prediction theory \cite{David_link_prediction},
we select the candidate vertex $w_{max}$ which guarantees that $nmf'(w_{max},u)+nmf'(w_{max},v)$ is maximum, where $nmf'(w,x)$ is defined in Eq.\ref{formlar3}.
\begin{equation}\label{formlar3}
 nmf'(w,x)=\left\{ \begin{array}{lll}
               0  &       &(x,w)\in E'\\
               nmf(w,x)&  &otherwise
              \end{array}
       \right.
\end{equation}
Where $x\in\{u,v\}$. This is referred to as the \emph{maximum mutual friend criterion} for adding edges. The more mutual friends between the two vertices, the less impact the edge addition will have on the utility of the graph.

The selection criteria described in the Eq.\ref{formlar3} only can be used for the candidates in the $1$-hop and $2$-hop neighbors.  For all the candidates $w$ in the $i$-hop neighbors with $i\geq 3$, the NMF of $(x,w)$ is $0$. In this situation, we randomly select a candidate vertex $w_{max}$ from $CV_i$.

As we anonymize edges in descending order of NMF, we must consider the different situations on the NMF of the new edge $(x,w_{max})$.
In the situation $nmf(x,w_{max})>=g_f$, if $nmf(x,w_{max})$ is not equal to any $g^{'}_f\in G_f$, $(x, w_{max})$ cannot be added into the graph. This is because we cannot anonymize this edge in descending order anonymization.
Otherwise,
we add $(x, w_{max})$ and mark it as ``anonymized". We put this edge into the group with NMF equal to $g^{'}_f$. 
If $nmf(x,w_{max})<g_f$, add $(x,w_{max})$ and mark it as ``unanonymized".

\subsubsection{Candidates dynamic removal}
After a new triangle was created with the vertex $w_{max}\in CV_i$, we need to consider the effect of this triangle on the other candidate vertices in $CV_i$. To ensure the linking $u$ or $v$ with vertices in $CV_i$ follows the ATPP, some vertices will be dynamically removed from $CV_i$.

If $w\in CV_i$ connected with the selected vertex $w_{max}$ and the edge $(w,w_{max})$ is anonymized, then we remove $w$ from $CV_i$. This is because adding either $(w,u)$ or $(w,v)$ creates a new triangle containing $(w,w_{max})$, and destroys the anonymization of $(w,w_{max})$.

For any vertex $w\in CV_i$ with $(w,w_{max})$ is unanonymized, if $(w_{max},x)$ is anonymized and $(w,x)\not\in E'$, then we remove $w$ from $CV_i$. This is because if we select this $w$ as a new maximum vertex, we need to add $(w,x)$ to create a triangle containing $(u,v)$, meanwhile created a triangle containing $(w_{max},x)$. This destroyed the anonymization of the edge $(w_{max},x)$.

\subsubsection{Edge anonymization}
From the nodes $u$ and $v$, \emph{BFS-based Edge Anonymization} traverses the graph in a breadth-first manner. The \emph{BFSEA} iteratively generates a candidate set $CV_i$ from the $i$-hop neighbors of $u$ and $v$, where $i$ increases from $1$ to $\infty$.
After getting the candidate set $CV_i$, \emph{BFSEA} iteratively selects the best one from $CV_i$ by \emph{candidates selection} and creates a triangle to increase the NMF of $(u,v)$, then updates the $CV_i$ by the \emph{candidates dynamic removal}.
These operations will break when the NMF of $(u,v)$ reaches the current group NMF $g_f$ or no more candidate vertex can be found from the whole graph.

If $nmf(u,v)$ reaches the current $g_f$, i.e. $(u,v)$ is anonymized successfully, we mark it as "anonymized".
If the \emph{BFSEA} cannot successfully anonymize this edge, adding new vertices can achieve the task. Linking one new vertex with the end vertices of this edge can increase the NMF of this edge by 1. The newly added edges have only one mutual friend, and will be anonymized at the last step of the anonymization algorithm.
The above scenario is a pathological case that rarely occurs as in our experiments, no new vertices were added in all cases.

By the breadth-first manner, the \emph{BFSEA} first link $u$ or $v$ with $w$ from the $1$-hop neighbors.
Thus after $(x, w)$ is added, the shortest path length (SPL) between $x\in\{u,v\}$ and $w$ will only decrease to 1 from 2 with little effect to the utility. Then we gradually increase the value of $i$, and link $u$ and $v$ with $w$ from the $i$-hop neighbors, which decreases the SPL between $x$ and $w$ from $i$ to $1$. Hence, we prefer the candidates from $i$-hop neighbors with smaller $i$ value, i.e. breadth-first manner, which can have less effect to the utility of the graph.




To get the $neig_i(u)$ and $neig_i(v)$ for every $i$, we execute the \emph{Breadth-First Search} with the time complexity as $O(|V|+|E|)$. When $i=1$, we need to compute the $neig_i(u)\cap neig_i(v)$ to ensure $(w,u,v)\neq\triangle$ stated in the \emph{candidates generation}, and the time complexity is $O(|V|)$. When $i\leq 2$, to get the best candidate from $CV_i$, we compute the $nmf(w,x)$, $x\in\{u,v\}$, with the time complexity as $O(|V|)$.
Hence, for each candidate set, the total running time of the NMF computation is $O(|V|^2)$.
When $i\geq 3$, we randomly select a candidate from $CV_i$ to create a triangle, and the time complexity is $O(1)$. Hence, the time complexity of \emph{candidates selection} is $O(|V|^2)$. Therefore, the time complexity of \emph{BFSEA} is $O(|V|^2)$.

As there are $O(|E|)$ edges need to be anonymized, the time complexity of the ADD algorithm is $O(|E||V|^2)$.

\vspace{-0.2cm}
\subsection{Algorithm ADD\&DEL}\label{sub:adddel}
Usually, anonymization combining edge deletion with addition will remove or add fewer edges than only applying edge addition. Indeed, it can improve the utility of the anonymized graph.
Before introducing the ADD\&DEL algorithm, we discuss the method on how to anonymize an edge by edge deletion.

\noindent\textbf{Edge-deletion.}
For an unanonymized edge $(u,v)$, the algorithm finds any candidate edge $(x,w)$, where $x$ is $u$ or $v$, which satisfies the following conditions.

\renewcommand{\theenumi}{\alph{enumi}}
\begin{enumerate}\vspace{-0.2cm}
  \item Both $(u,w)$ and $(v,w)$ exist and are unanonymized.
  \item For any vertex $z$ linked with $x$ and $w$, edges $(x,z)$ and $(w,z)$ are still unanonymized.
  \item If both $(u,w)$ and $(v,w)$ satisfy condition b), we choose the one with fewer mutual friends.\vspace{-0.2cm}
\end{enumerate}

Condition c) is the reverse of the maximum mutual friend criterion for adding edge. The fewer the mutual friends, the weaker the relationship. Hence dropping the edge has less impact to the utility.
After $(x, w)$ is deleted, the shortest path length between $x$ and $w$ will only increase to 2 from 1 with little effect to the utility. Condition a) and b)
follows the anonymized triangle preservation principle to guarantee that there will be no effect on the already anonymized edges.

For an unanonymized edge $(u,v)$, \emph{edge-deletion} initially finds all candidate edges satisfying the edge-deletion conditions, and then puts them into the set $CE$. During each iteration, the edge $e_{min}\in CE$ with the least mutual friends will be removed from the graph and the set $CE$. The algorithm stops when the NMF of $(u,v)$ reaches the group NMF $g_f$ or $CE$ becomes an empty set. If $CE$ is empty and the NMF of $(u,v)$ is not equal to $g_f$, the anonymization of $(u,v)$ is unsuccessful; Otherwise, we successfully anonymize this edge and mark it as ``anonymized".

The \emph{edge-deletion} is the reverse of the methods in ADD algorithms. The running time mainly costs on the computing of mutual friends, so the complexity of \emph{edge-deletion} is $O(|V|^2)$.

\noindent\textbf{The ADD\&DEL Algorithm.} This algorithm is shown in Algorithm 2, which anonymizes the graph by edge addition and deletion. Similar to the ADD algorithm, ADD\&DEL checks the number of unanonymized edges with NMF equal to the NMF of the first unanonymized edge in sorted sequence $\bm{l}^{(u)}$. If there are more than $k$ edges, we put them into this group and start another group. Otherwise, we need to anonymize edges to form this group.
To gradually anonymize edges and create this group, we initially set the group NMF, $g_f$, as the mean value of NMFs of the first $k$ unanonymized edges. We record \emph{all initial information} before anonymizing this group. For the unanonymized edge with NMF greater than $g_f$, we use edge-deletion to anonymize it. If we cannot successfully anonymize this edge, we set $g_f=g_f+1$ and roll back to \emph{all initial information}. For the unanonymized edge with NMF less than $g_f$, we apply the ADD algorithm to anonymize it. We gradually anonymize unanonymized edges in sorted sequence $\bm{l}^{(u)}$ until this group has $k$ edges, and start another group.

In the ADD\&DEL algorithm, an edge will be anonymized by either Edge-deletion or methods of the ADD algorithm. Therefore, the time complexity of anonymizing an edge is $O(|V|^2)$, and the time complexity of the ADD\&DEL algorithm is $O(|E||V|^2)$.

\begin{figure}[!t]
\centering
\includegraphics[width=0.48\textwidth]{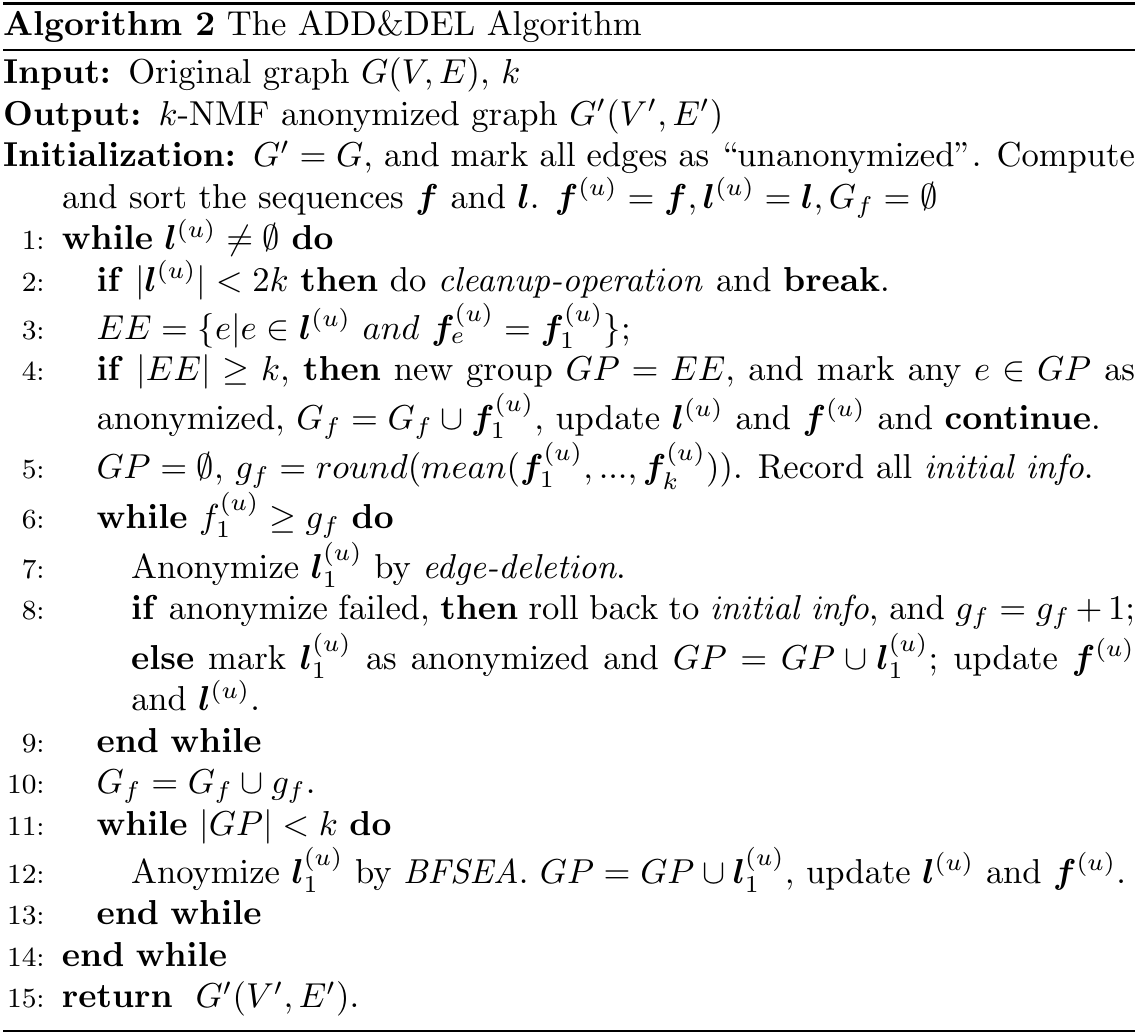}
\label{alg:AddDel}
\vspace{-0.5cm}
\end{figure}

\vspace{-0.2cm}
\subsection{$k_1$-degree Anonymization Based on $k_2$-NMF Anonymization}\label{subsec:KDA-ALGO}
In this subsection, we propose the KDA algorithm on anonymizing the $k_2$-NMF anonymized graph $G'$ to satisfy $k_1$-degree anonymity. To maintain the $k_2$-NMF anonymity of $G'$, the KDA algorithm does not change the NMF of edges in $G'$ when performing anonymization. Proposition \ref{prop3} stated that the NMF of an edge is related on the number of triangles in which this edge participate, so we anonymize the graph $G'$ without adding new triangles, i.e., the anonymized triangle preservation principle. We can connect two vertices with shortest path length (SPL) no less than three to guarantee that no new triangles will be introduced. Then the NMF of newly added edge is zero. 
As the degree distribution of the social network follows the power law \cite{Faloutsos:power_law_degrees}, we only need to anonymize these vertices with large degrees.

The $KDA$ algorithm is similar to the ADD algorithm.
The unanonymized vertices are sorted in descending order of their degrees. We gradually group and anonymize them only by edge addition. The vertices in the same group have same degree. To start a new group, $KDA$ set the group degree $g_d$ as the greatest degree of unanonymized vertices. If there are less than $k$ vertices in this group, we anonymize the unanonymized vertices in descending order of their degrees. If this group has more than $k$ vertices, we compute the $C_{merge}$ and $C_{new}$ for the next unanonymized vertex, and decide whether put it into this group or start a new group.

Suppose that the $i$-hop neighbors of vertex $u$ is $neig_i(u)$. To anonymize the unanonymized vertex $u$, $KDA$ iteratively and randomly select an unanonymized vertex $w_{max}$ from $neig_3(u)$ and connect $u$ and $w_{max}$.
If the vertex $u$ cannot be anonymized, $KDA$ update the $neig_3(u)$ based on the newly added edges and repeat the above process. If $u$ still cannot be anonymized, we select the candidate vertex from $neig_4(u), neig_5(u)$ and so on until $u$ is anonymized.

When anonymizing a vertex, the KDA algorithm searches the graph in a breadth-first manner to get the candidate vertices. In the worst case, the KDA searches the whole graph and the time complexity is $O(|E|+|V|)$. As there are $O(V)$ vertices needed to be anonymized, the time complexity of the KDA algorithm is $O(|E||V|+|V|^2)$ in the worst case.

\section{Experimental Results}
In this section, we conduct experiments on real data sets to evaluate the performance of the proposed graph anonymization algorithms.

\vspace{-0.2cm}
\subsection{Datasets}\label{datasets}
We conduct our experiments on three real datasets: ACM, Cora, and Brightkite. All datasets are preprocessed into simple undirected graphs without self-loop and multiple edges. We also remove the isolated vertices from the graph.

\textbf{ACM}: This dataset was extracted from ACM digital library. We extracted papers published in 12 conference proceedings on computer science before the year 2011.
We derive a graph describing the citations between papers. If one paper cites another paper, an undirected edge will connect both corresponding vertices. The graph includes 7,315 vertices and 16,203 edges.

\textbf{Cora}: This dataset is composed of a number of scientific papers on computer science \cite{A.McCallum:dataset}. We extract the collaborations between authors to derive the graph. If two authors had co-authored some papers
 they would be connected. After we removed the authors without any collaboration, the graph contains 14,076 vertices and 72,871 edges.

\textbf{Brightkite}: This dataset shows the friendships between users in the social network Brightkite over the period of April 2008 to October 2010.
The graph consists of 58,228 nodes and 214,078 edges, and is available at the SNAP \cite{Stanford_Large_Network}.

\vspace{-0.2cm}
\subsection{Mutual Friend Attack in Real Data}
In the $k$-degree anonymization model, the adversary re-identifies a vertex using the degree of this vertex. In the $k$-NMF anonymization model, the adversary re-identifies an edge using the NMF of this edge. We compare both attacks on the real datasets listed in Subsection \ref{datasets}, and show the results in Table \ref{table:violation}. We removed all labels in three datasets. From Table \ref{table:violation}, we can see that the number of edges violating $k$-NMF anonymity can be sizable when we set $k$ from 5 to 100. It is a very easy way for an adversary to take the mutual friend attack. $k$-NMF anonymization problem can be seen as a parallel of the $k$-degree anonymization problem.

\begin{table}[!t]
\footnotesize
\caption{The numbers of vertices violating $k$-degree anonymization and edges violating $k$-NMF anonymization}
\begin{tabular}{|c|c|c|c|c|c|c|}
\hline
~& \multicolumn{2}{|c|}{ACM} & \multicolumn{2}{|c|}{Cora} & \multicolumn{2}{|c|}{Brightkite} \\
\hline
$k$&$k$-deg&$k$-NMF&$k$-deg&$k$-NMF&$k$-deg&$k$-NMF\\
\hline
5& 54& 28& 141&  106 &  266 &  93 \\
\hline
10& 75&  28 & 267 & 179 & 533 & 129\\
\hline
15& 103 & 43 & 408 & 277 & 705& 285\\
\hline
20& 137 & 62 & 446& 349& 795& 393\\
\hline
25& 162 & 62 & 584 & 488&  891 & 598\\
\hline
30& 221 &  62  & 752 & 575 & 1088 & 762\\
\hline
50& 262 & 99 & 1142  & 733 & 1425 & 1297\\
\hline
100& 526 & 226 & 1472  & 1350 & 2326 & 2578\\
\hline
\end{tabular}
\label{table:violation}
\end{table}

\vspace{-0.2cm}
\subsection{Evaluating $k$-NMF Anonymization Algorithms}
We evaluate the performance of the Greedy and Intuitive ADD algorithms and the ADD\&DEL algorithm
by measuring the average clustering coefficient, average path length, betweenness centrality and the ratios of edges change.
Figures 5-8 show the results, where ADD-Int and ADD-Gre stand for the ADD algorithm with IntuitGroup and GreedyGroup respectively. ADD\&DEL stands for the ADD\&DEL algorithm.

\begin{figure*}[t]
\centering\vskip -0.2in
\subfigure[ACM]{\label{acm-cc}
\raisebox{-0.2cm}{\includegraphics[width=0.3\textwidth, height=1.0in]{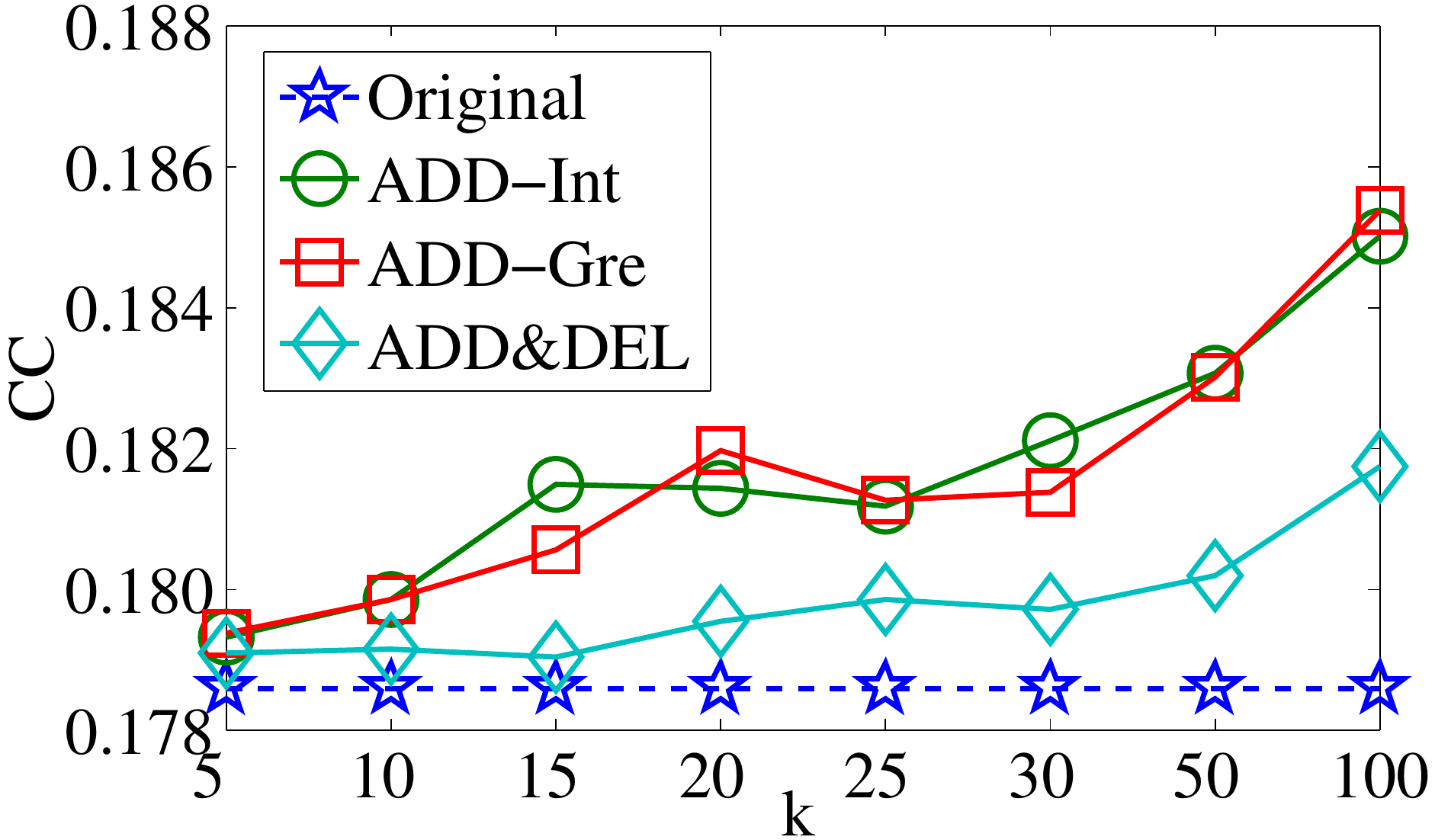}}}
\subfigure[Cora]{\label{core-cc}
\raisebox{-0.2cm}{\includegraphics[width=0.3\textwidth, height=1.0in]{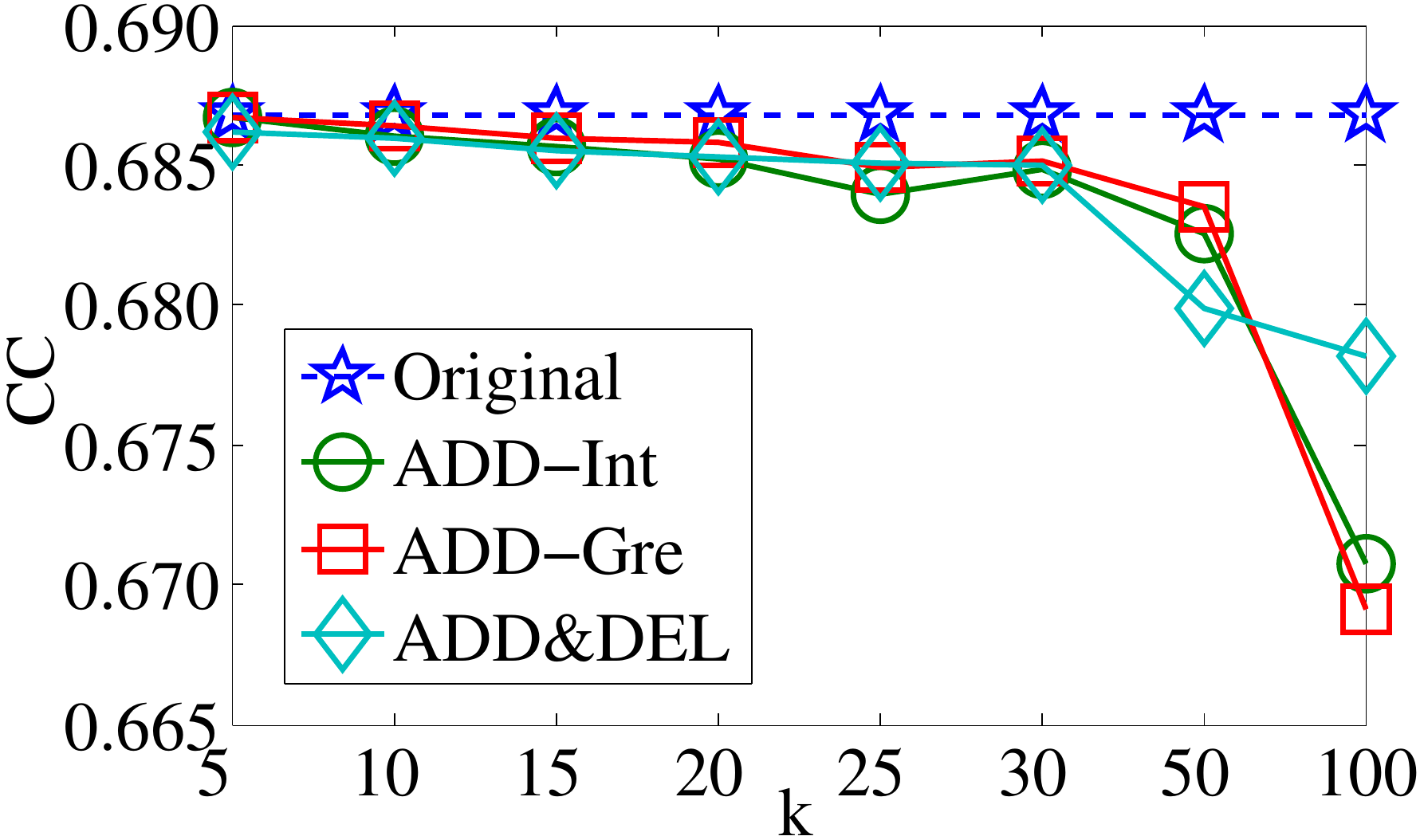}}}
\subfigure[Brightkite]{\label{brightkite-cc}
\raisebox{-0.2cm}{\includegraphics[width=0.3\textwidth, height=1.0in]{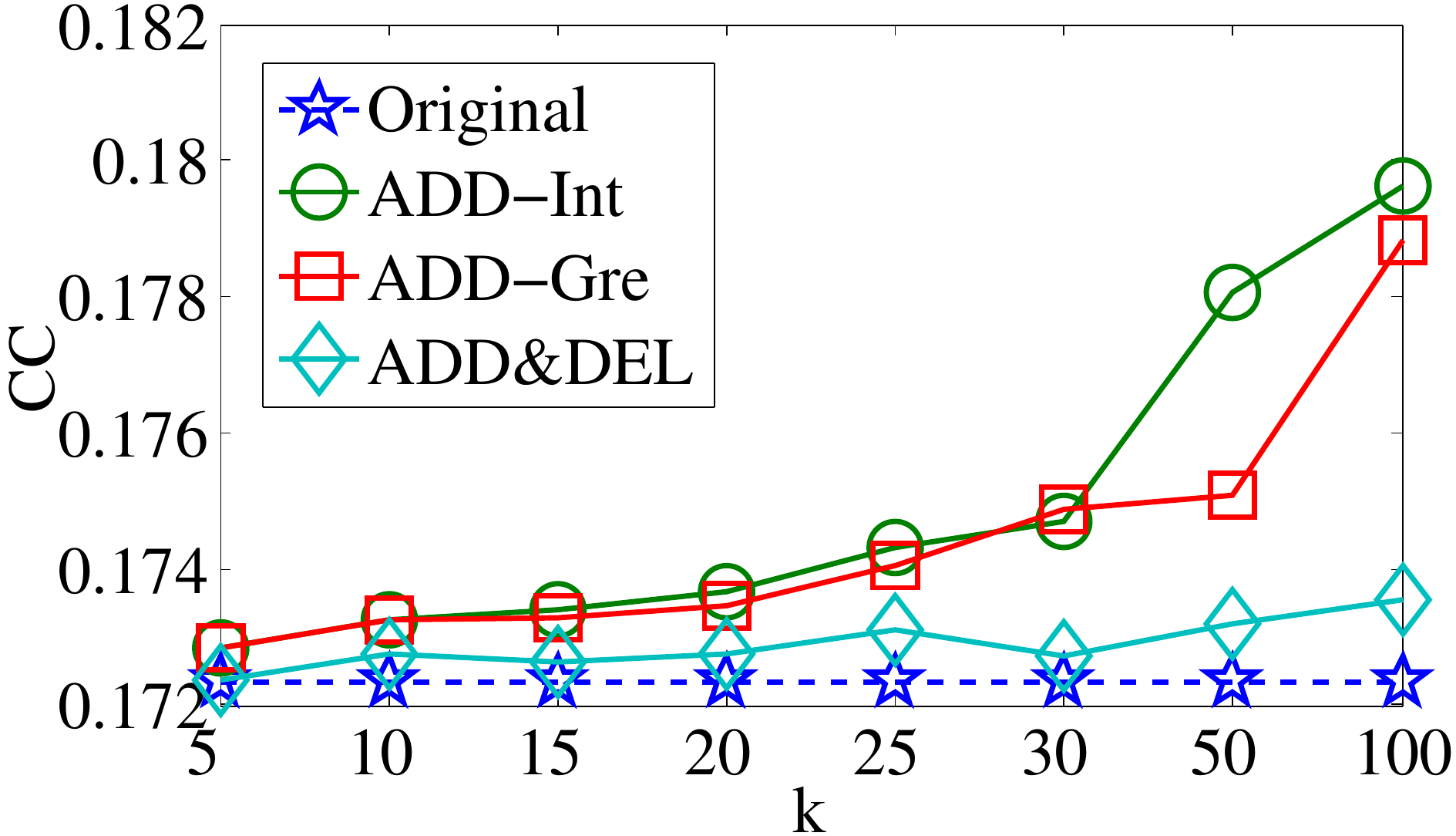}}}
\caption{Clustering coefficients}\vskip -0.1in
\label{cc}
\end{figure*}

\noindent\textbf{Average Clustering Coefficient (CC)}: We first compare the average clustering coefficients of the anonymized graphs with the original graph, and the results are shown in Figure \ref{cc}.
The CC values on datasets ACM and Brightkite increase when $k$ increases, but decreased on dataset Cora when $k$ increases. Hence no clear trend on CC change can be concluded, however the average clustering coefficients derived by our three methods deviate slightly from the original values on three datasets. The ADD\&DEL performs better than the two ADD algorithms in Figure \ref{cc}, and the ADD algorithm with GreedyGroup looks slightly better than the algorithm with IntuitGroup.

\begin{figure*}[]
\centering
\subfigure[ACM]{\label{acm-apl}
\raisebox{-0.2cm}{\includegraphics[width=0.3\textwidth, height=1.0in]{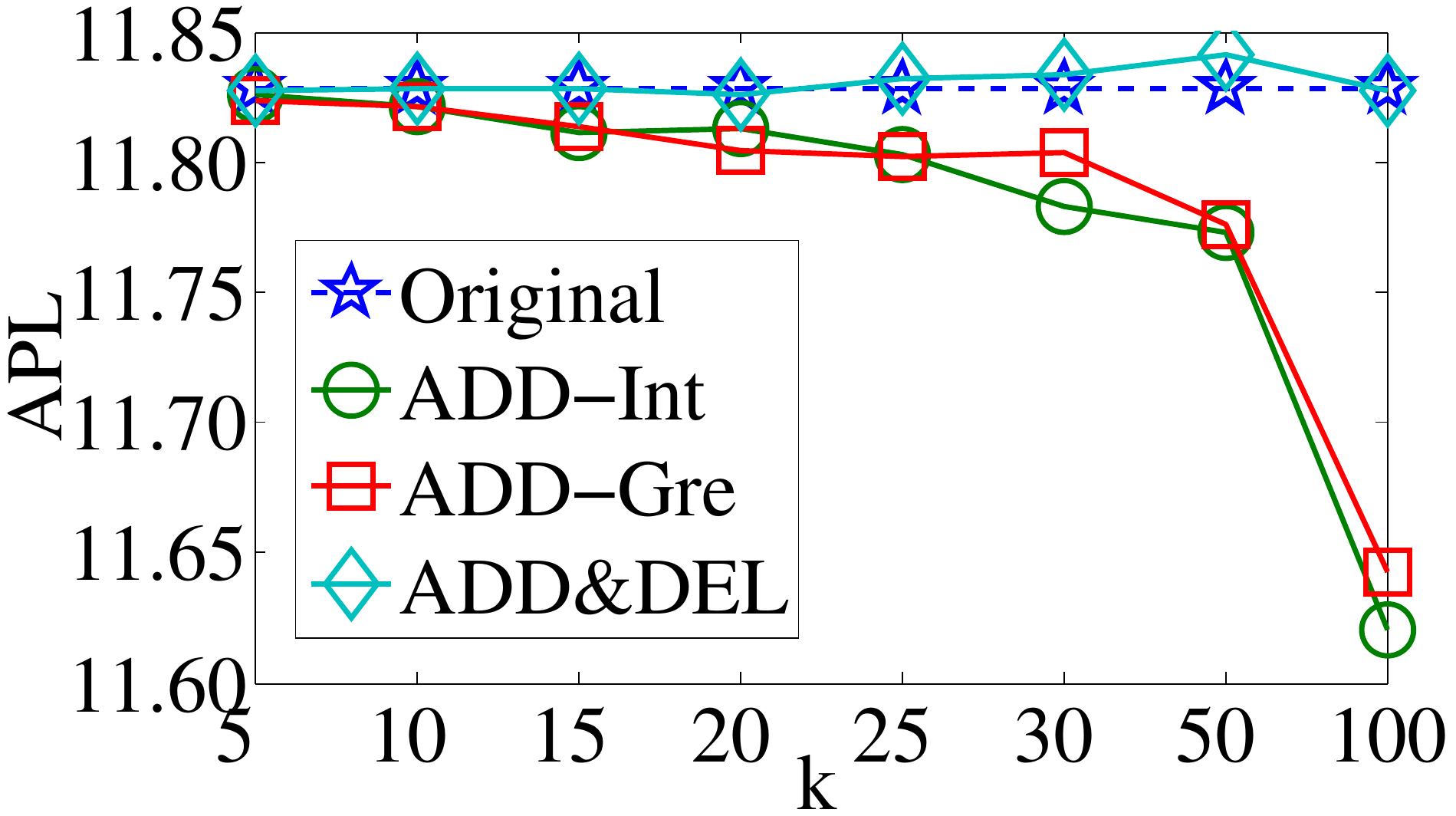}}}
\subfigure[Cora]{\label{core-apl}
\raisebox{-0.2cm}{\includegraphics[width=0.3\textwidth, height=1.0in]{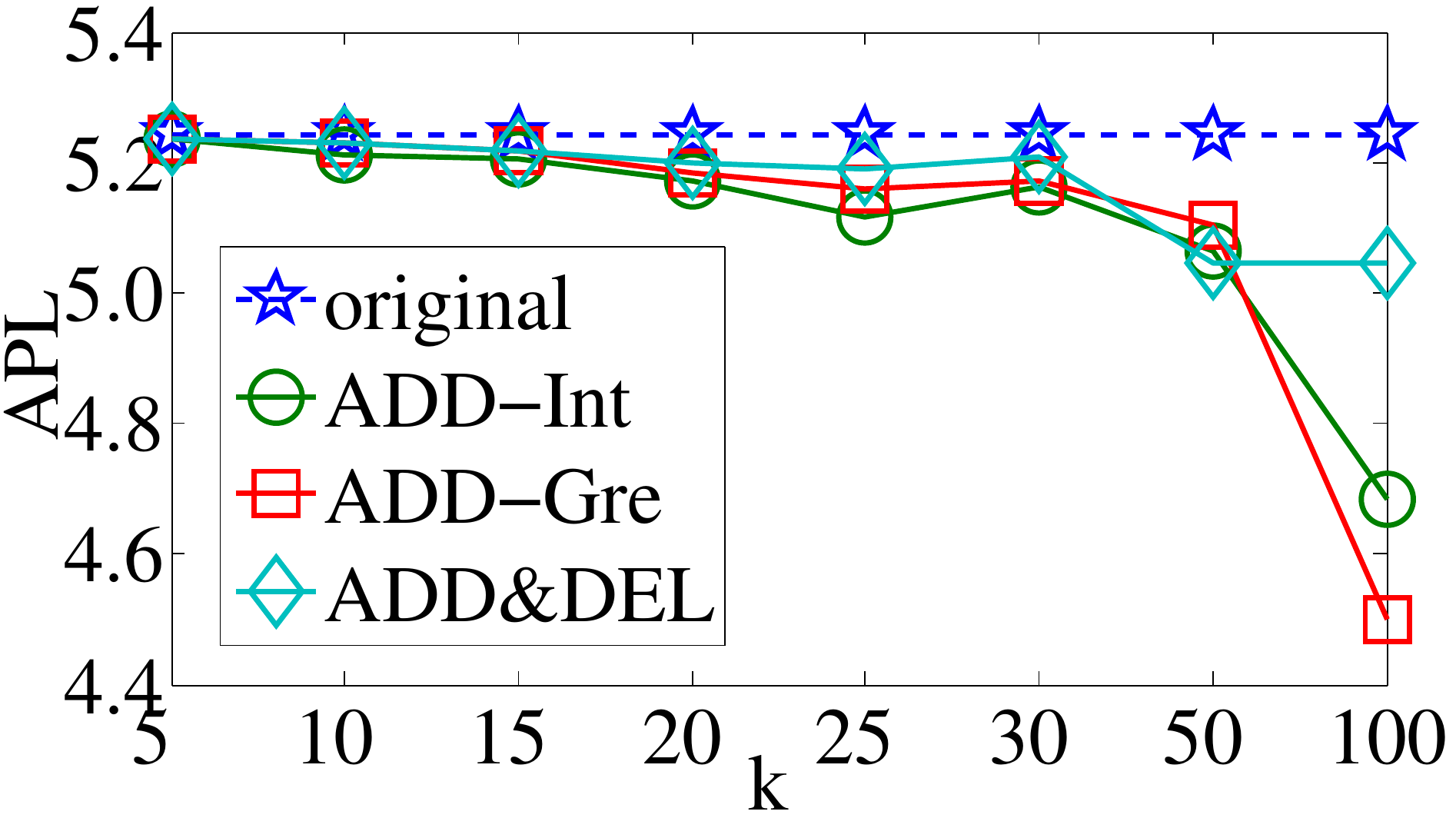}}}
\subfigure[Brightkite]{\label{brightkite-apl}
\raisebox{-0.2cm}{\includegraphics[width=0.3\textwidth, height=1.0in]{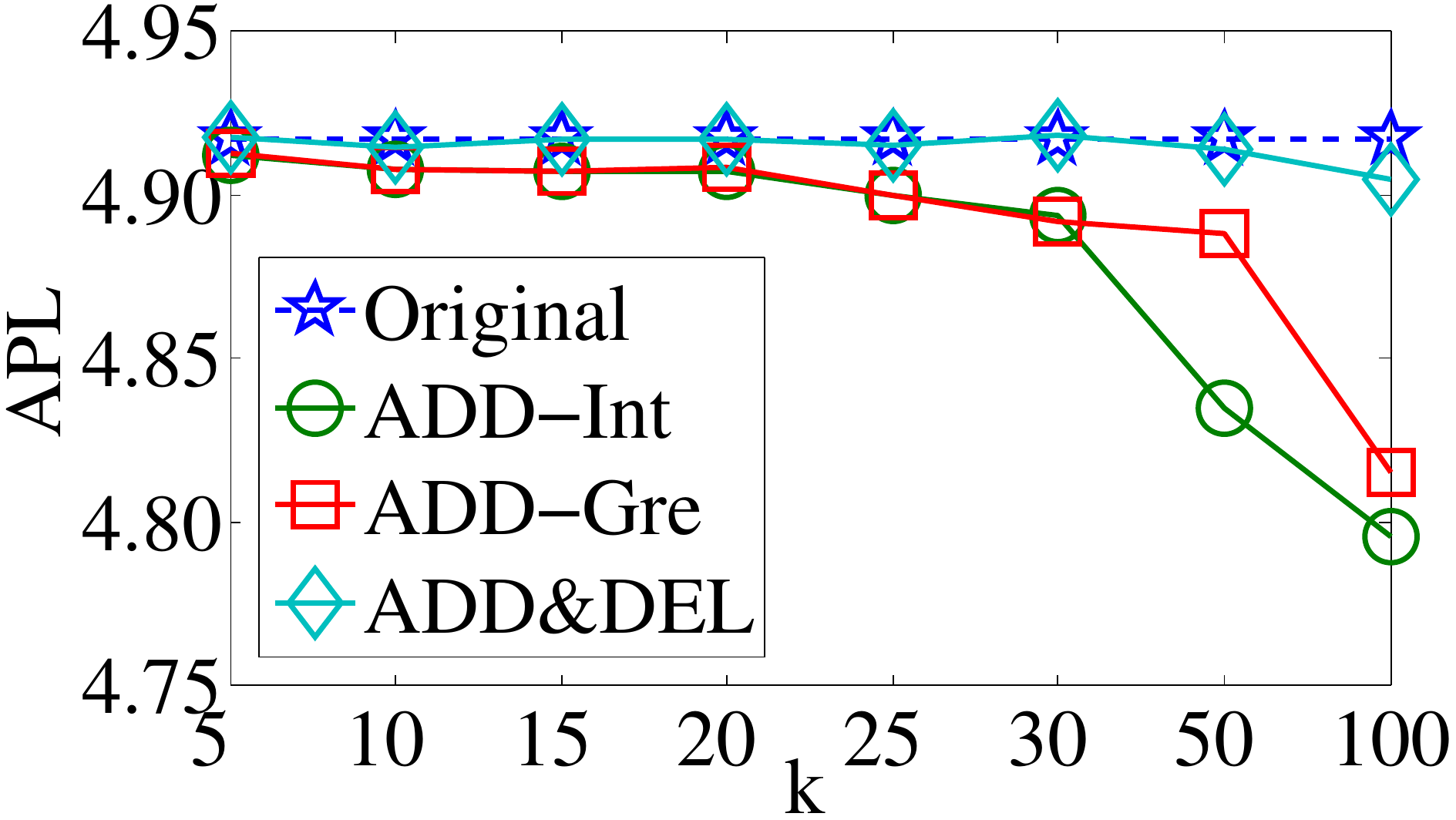}}}
\caption{Average path lengths}\vskip -0.1in
\label{apl}
\end{figure*}

\noindent\textbf{Average Path Length (APL)}: Figure \ref{apl} shows the average path lengths for the anonymized graphs and the original graphs on three datasets.
The APL of the graph anonymized by the ADD\&DEL algorithm is very close to the APL of the original graph. By adding and deleting edges, the ADD\&DEL algorithm can preserve more utility than the ADD algorithm. Besides, the differences of APL between the graphs anonymized by our methods and the original graphs are very small, and the largest difference value is 0.8 when $k$ is set as 100 on the dataset Cora.

\begin{figure*}[]
\centering
\subfigure[ADD\&DEL algorithm]{\label{core_bc_adddel}
\raisebox{-0.2cm}{\includegraphics[width=0.3\textwidth, height=1.0in]{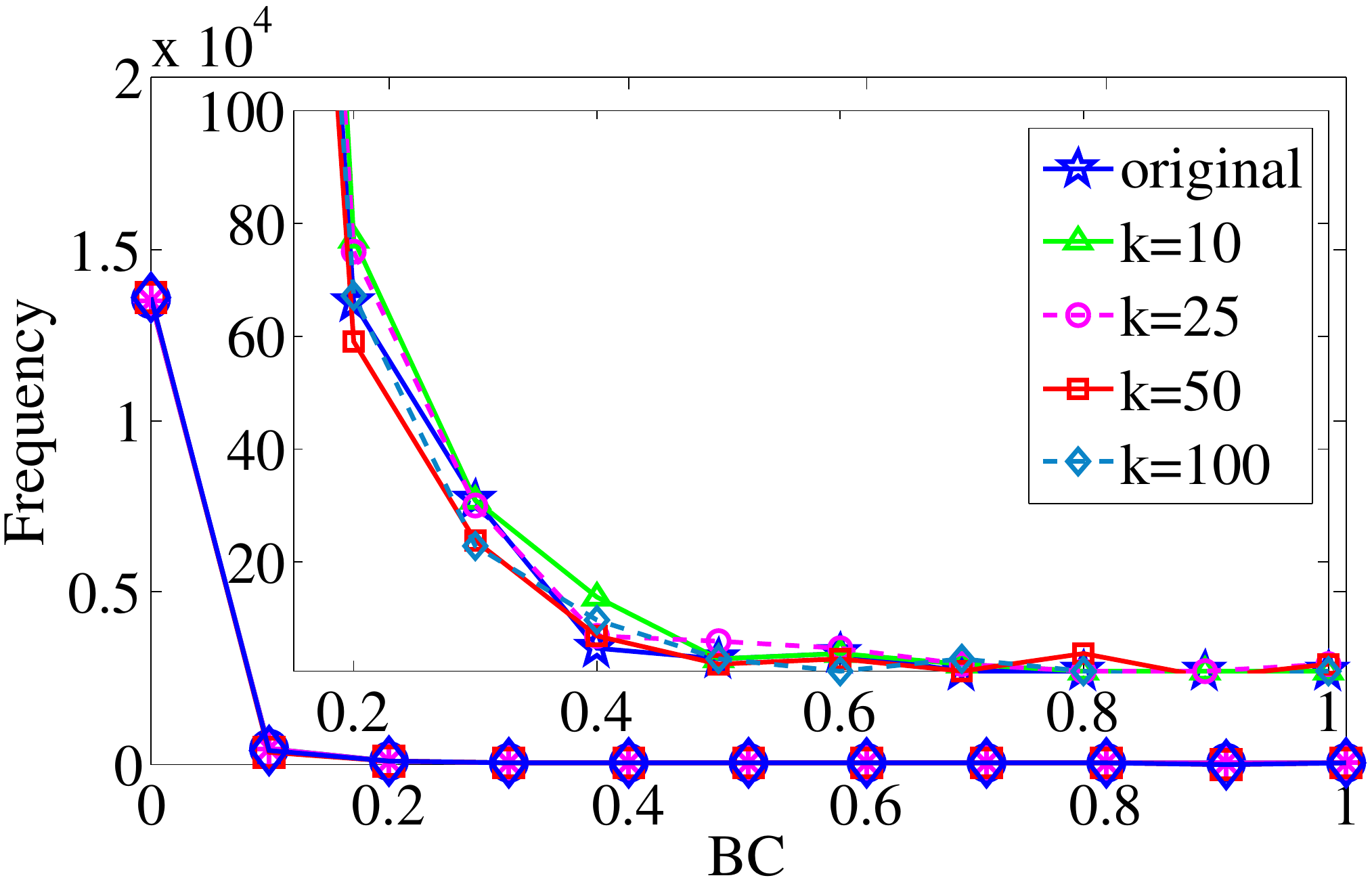}}}
\subfigure[ADD algorithm with GreedyGroup]{\label{core_bc_addGre}
\raisebox{-0.2cm}{\includegraphics[width=0.3\textwidth, height=1.0in]{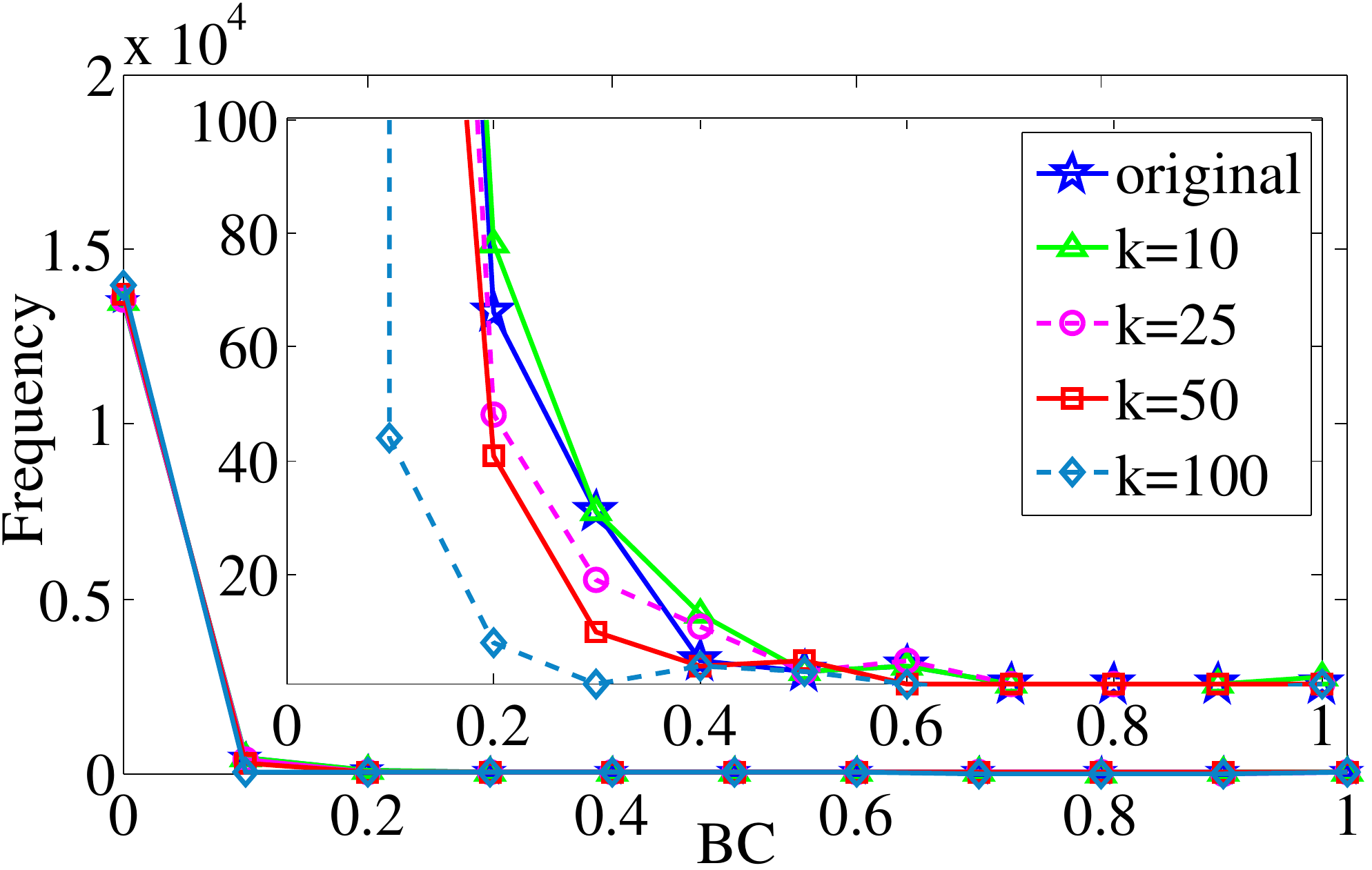}}}
\subfigure[$k$ is 25]{\label{core_bc_k_25}
\raisebox{-0.2cm}{\includegraphics[width=0.3\textwidth, height=1.0in]{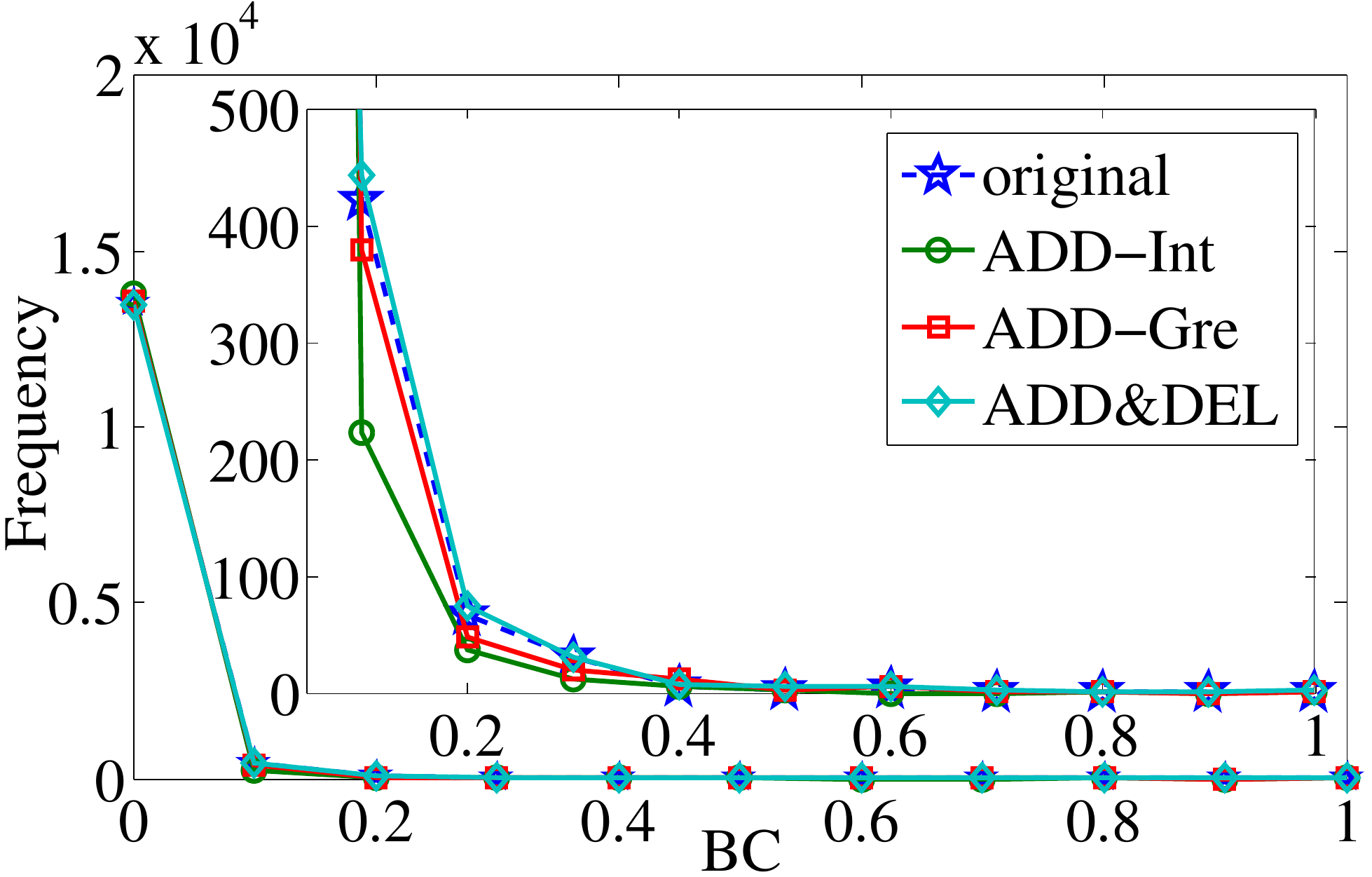}}}
\caption{Betweenness centrality distributions on Cora}\vskip -0.1in
\label{bc}
\end{figure*}

\noindent\textbf{Betweenness Centrality (BC)}: All the plots of the average betweenness centralities are very similar to the plots of the \emph{APL}. Hence we show the distribution of betweenness centralities of all vertices in Figure \ref{bc}. Due to space constraints, we only show the results on Cora. The sub-figures in Figures \ref{core_bc_adddel}, \ref{core_bc_addGre} and \ref{core_bc_k_25} enlarge the details on the frequency varied from 0 to 100. Clearly, in Figures \ref{core_bc_adddel} and \ref{core_bc_addGre}, ADD\&DEL performs better than the ADD algorithm with GreedyGroup, and shows little sensitivity to the value of $k$ while ADD with GreedyGroup degrades as $k$ increases. Also Figure \ref{core_bc_k_25} shows that ADD\&DEL performs better than the ADD algorithms.

\begin{figure*}[]
\centering
\subfigure[ACM]{\label{acm_edge_change}
\raisebox{-0.2cm}{\includegraphics[width=0.3\textwidth, height=1.0in]{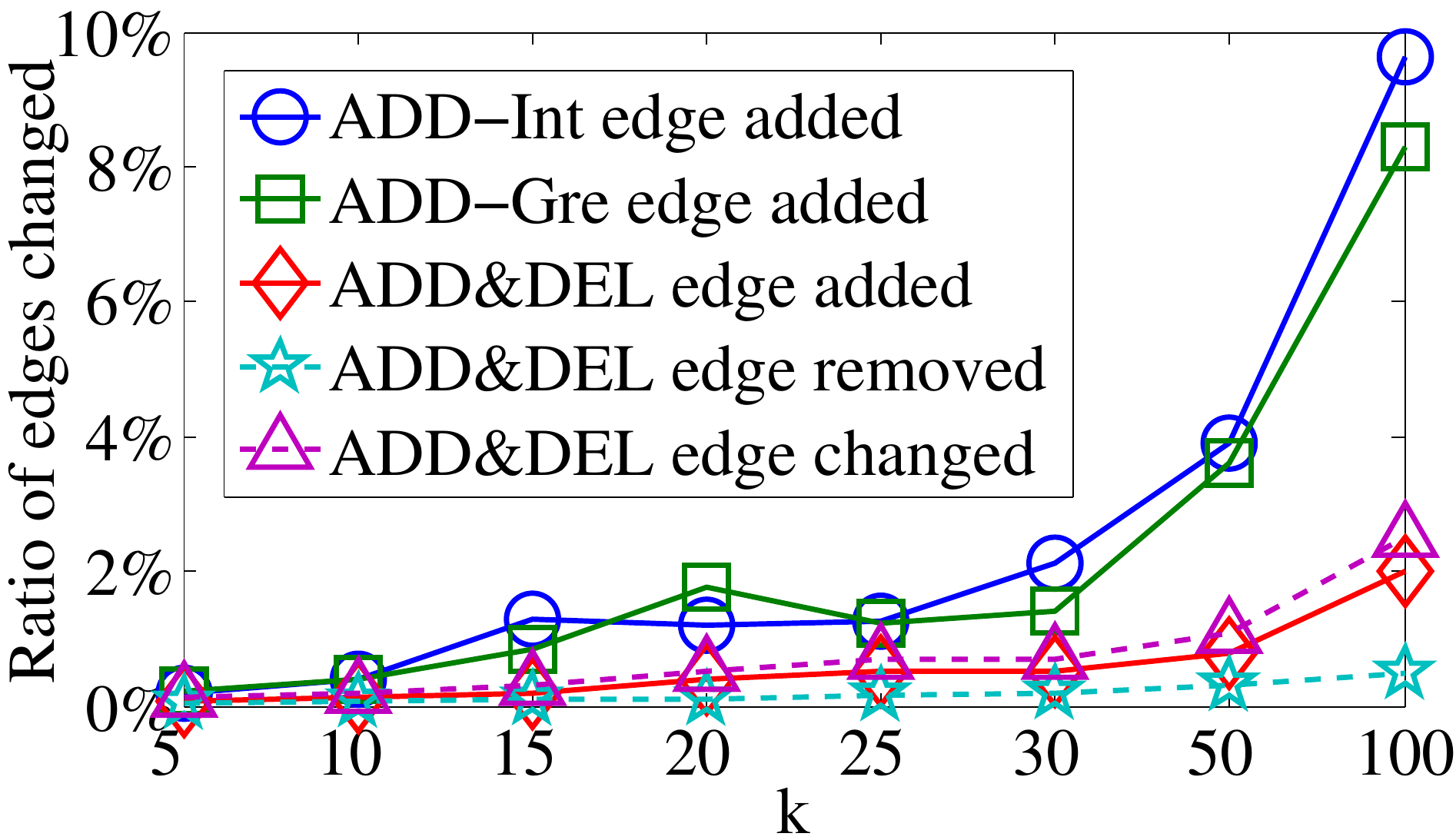}}}
\subfigure[Brightkite]{\label{brightkite_edge_change}
\raisebox{-0.2cm}{\includegraphics[width=0.3\textwidth, height=1.0in]{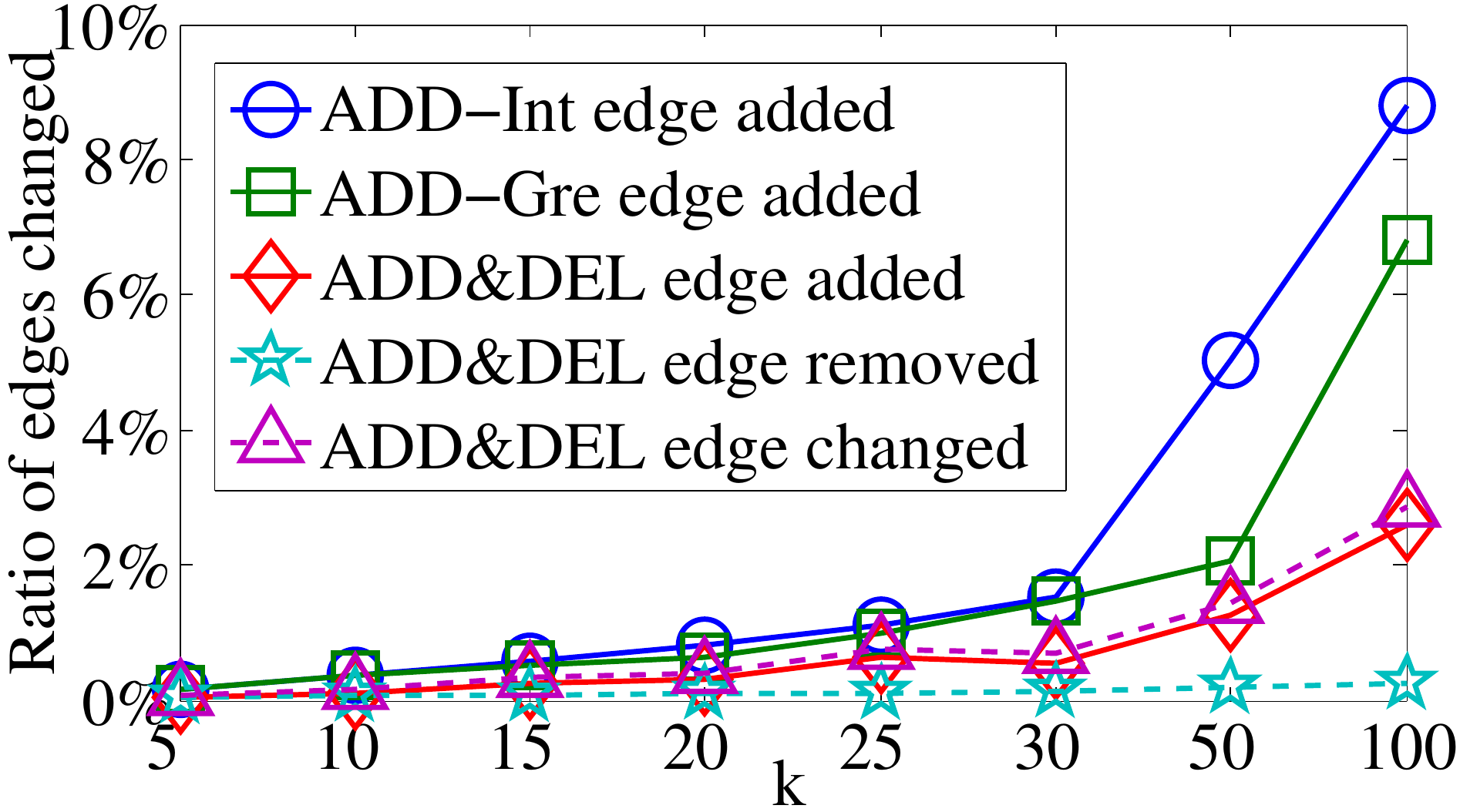}}}
\subfigure[ACM]{\label{acm_vertex_change}
\raisebox{-0.2cm}{\includegraphics[width=0.3\textwidth, height=1.0in]{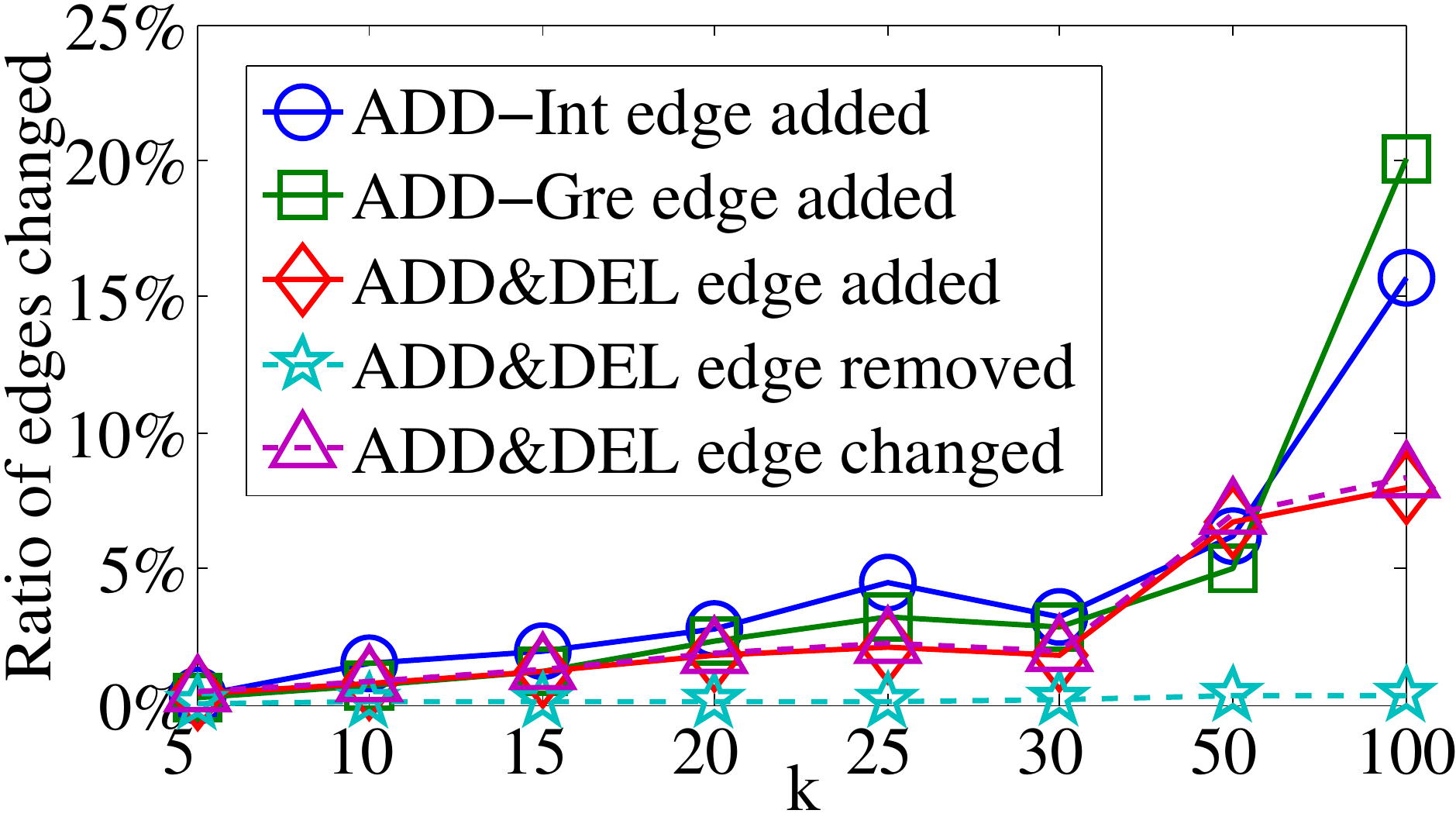}}}
\caption{Edge changes}\vskip -0.1in
\label{edge_change}
\end{figure*}

\noindent\textbf{Percentages of edges changed}:
As there is no vertex addition occurred in all cases considered under ADD and ADD\&DEL which do not perform node deletion operations, we consider the edge changes.
Figure \ref{edge_change} shows the edge changes on the original graphs. The changes on ADD\&DEL includes the ratios of edges added and removed.
The ADD\&DEL algorithm changed fewest edges, and the ADD algorithm with GreedyGroup added fewer edges than the algorithm with IntuitGroup.

From the above evaluation, we can see that our algorithms can preserve the utility of the original graph effectively. Among them, ADD\&DEL performs better than the ADD algorithm, and GreedyGroup performs better than IntuitGroup.

\vspace{-0.2cm}
\subsection{Evaluating the $KDA$ Algorithm}

In this subsection, we evaluate the performance of the KDA algorithm in Section \ref{subsec:KDA-ALGO}
 , and compare it with the classic $k$-degree anonymization algorithm in \cite{K.Liu:k-anonymization}.

\begin{figure*}[]
\centering
\subfigure[CC]{\label{acm-deg-cc}
\raisebox{-0.2cm}{\includegraphics[width=0.3\textwidth, height=1.0in]{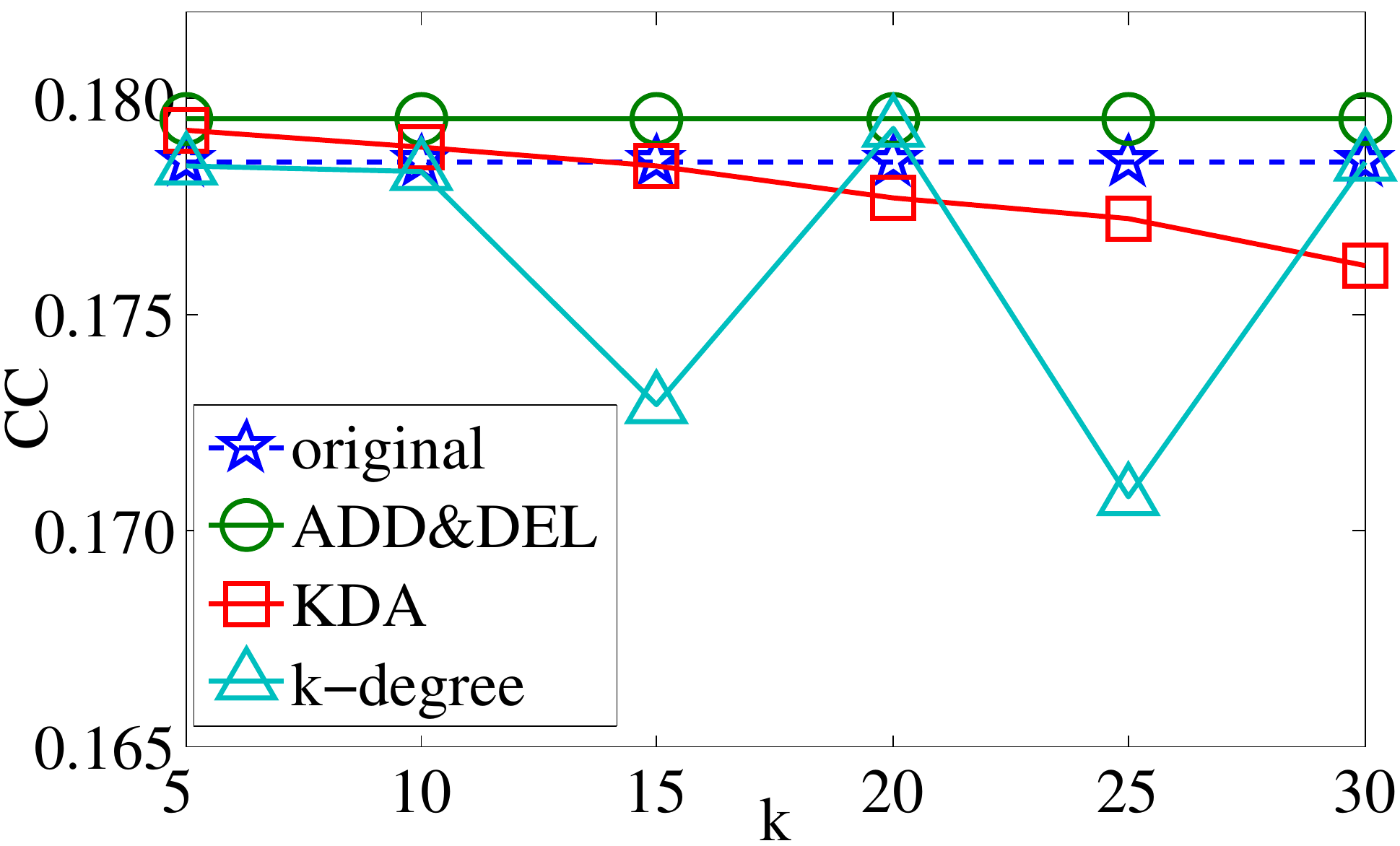}}}
\subfigure[APL]{\label{acm-deg-apl}
\raisebox{-0.2cm}{\includegraphics[width=0.3\textwidth, height=1.0in]{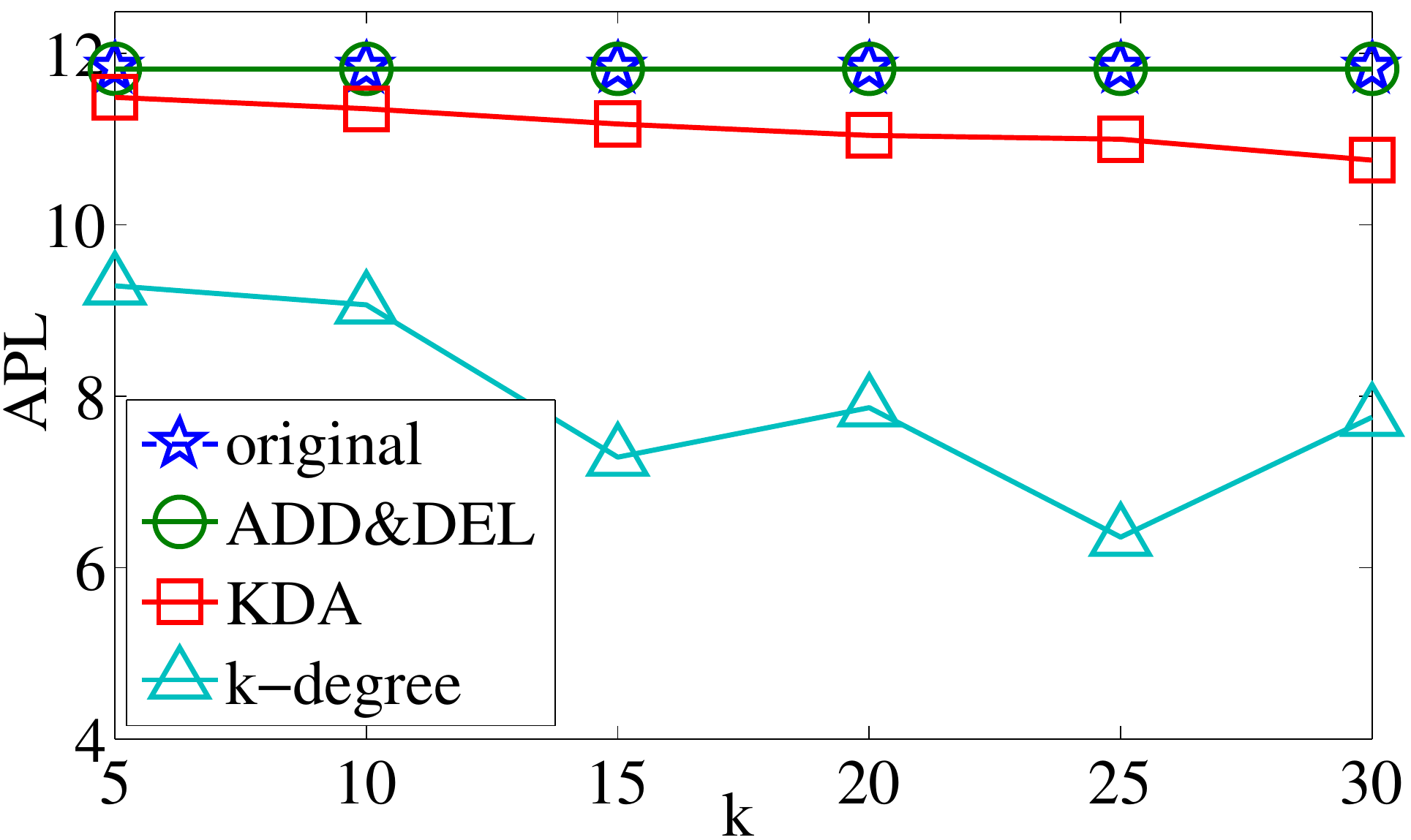}}}
\subfigure[BC]{\label{acm-deg-bc}
\raisebox{-0.2cm}{\includegraphics[width=0.3\textwidth, height=1.1in]{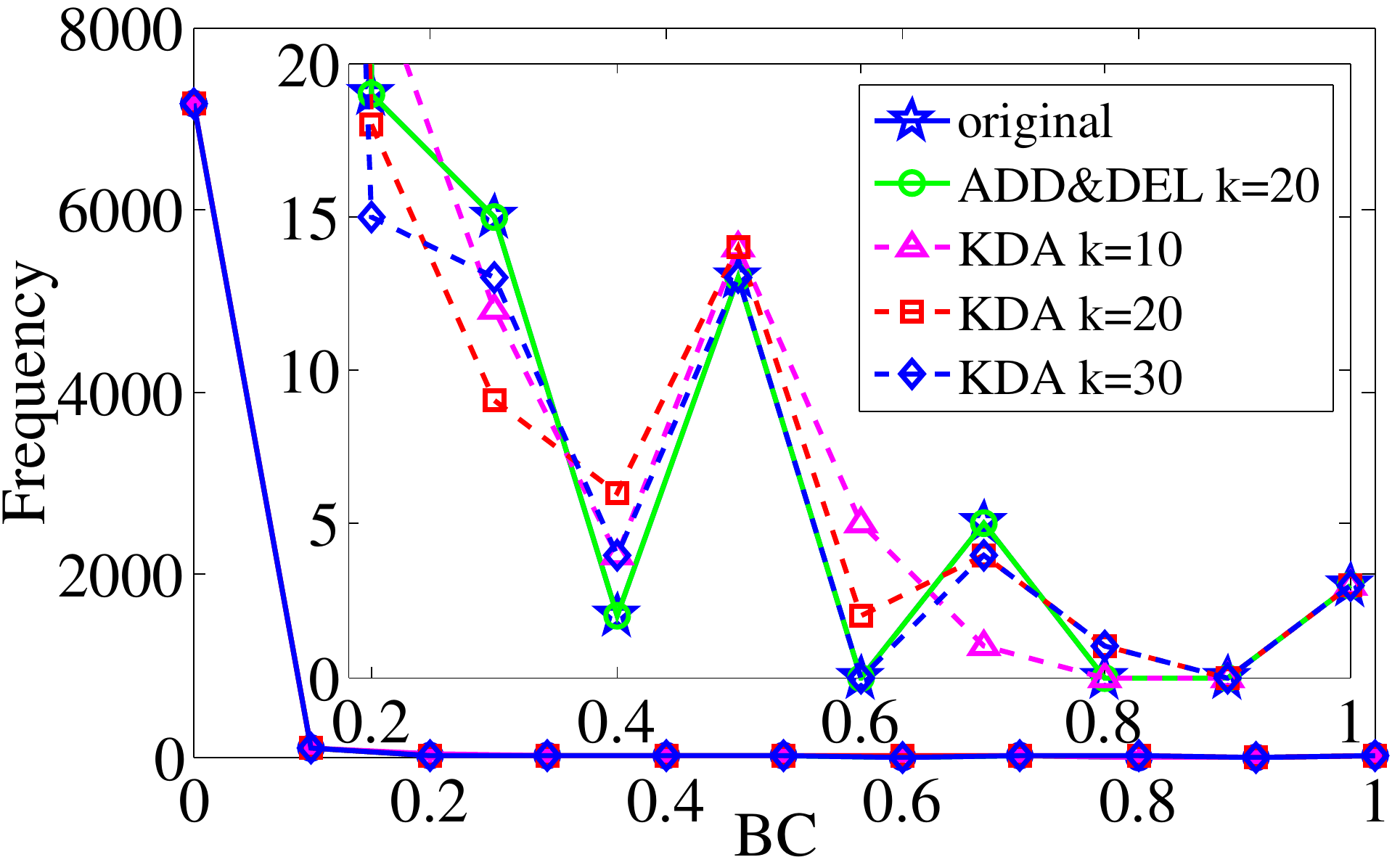}}}
\caption{k-degree anonymization on 20-NMF anonymized graph of ACM }\vskip -0.1in
\label{acm-deg}
\end{figure*}

\begin{figure*}[]
\centering
\subfigure[CC]{\label{cora-deg-cc}
\raisebox{-0.2cm}{\includegraphics[width=0.3\textwidth, height=1.0in]{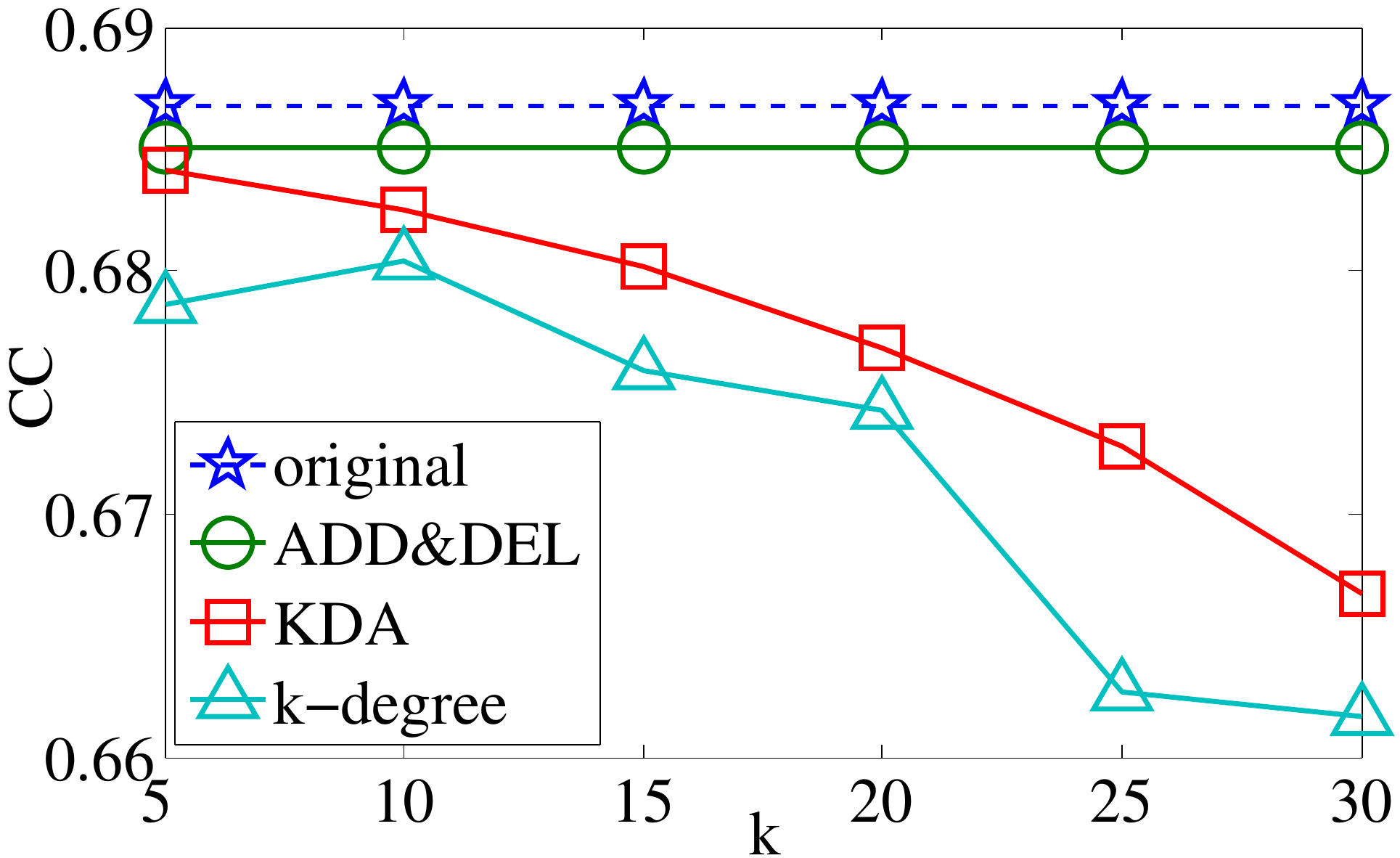}}}
\subfigure[APL]{\label{cora-deg-apl}
\raisebox{-0.2cm}{\includegraphics[width=0.3\textwidth, height=1.0in]{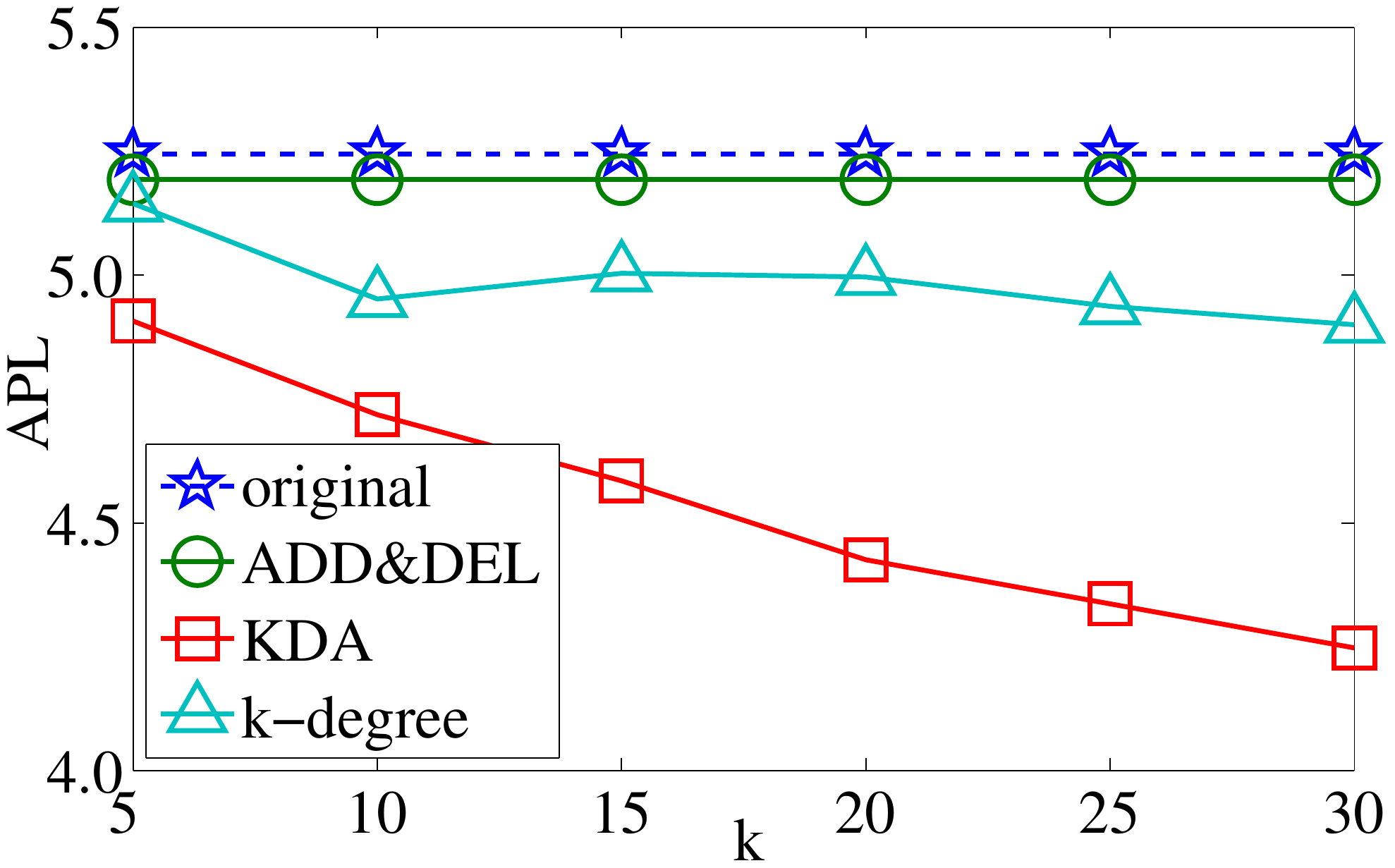}}}
\subfigure[BC]{\label{cora-deg-bc}
\raisebox{-0.2cm}{\includegraphics[width=0.3\textwidth, height=1.1in]{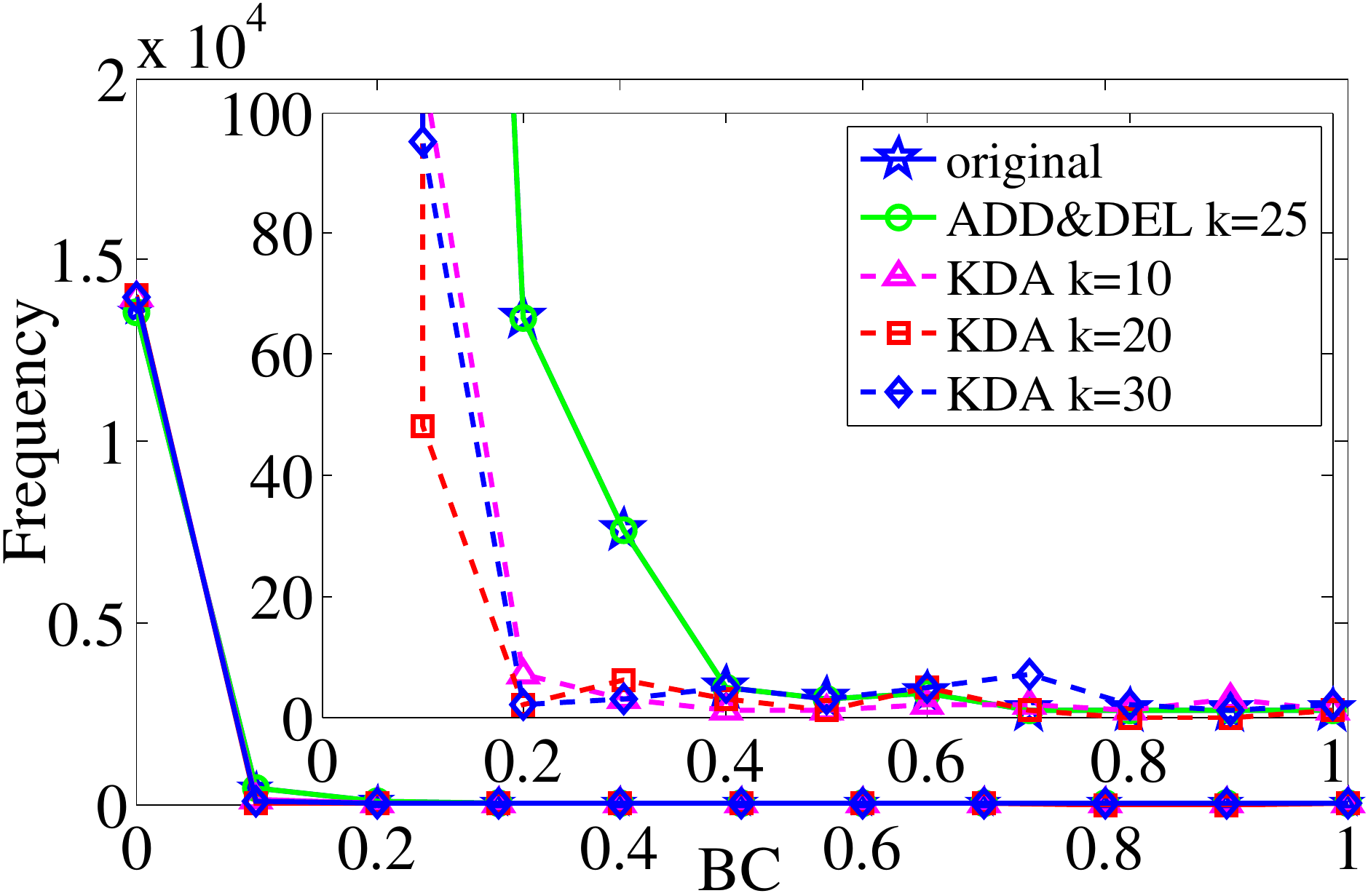}}}
\caption{k-degree anonymization on 25-NMF anonymized graph of Cora}\vskip -0.1in
\label{cora-deg}
\end{figure*}

\begin{figure*}[]
\centering
\subfigure[CC]{\label{brightkite-deg-cc}
\raisebox{-0.2cm}{\includegraphics[width=0.3\textwidth, height=1.0in]{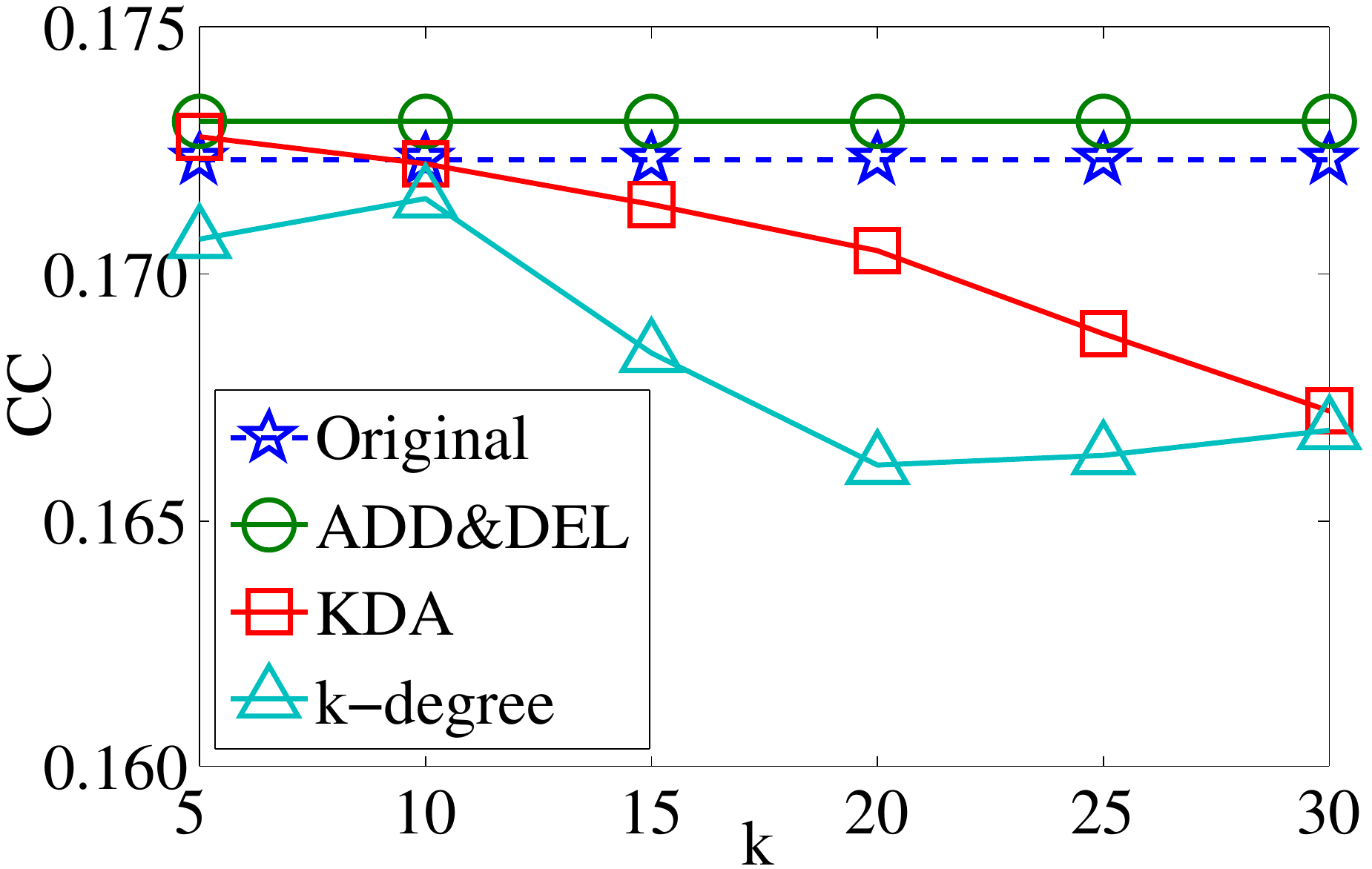}}}
\subfigure[APL]{\label{brightkite-deg-apl}
\raisebox{-0.2cm}{\includegraphics[width=0.3\textwidth, height=1.0in]{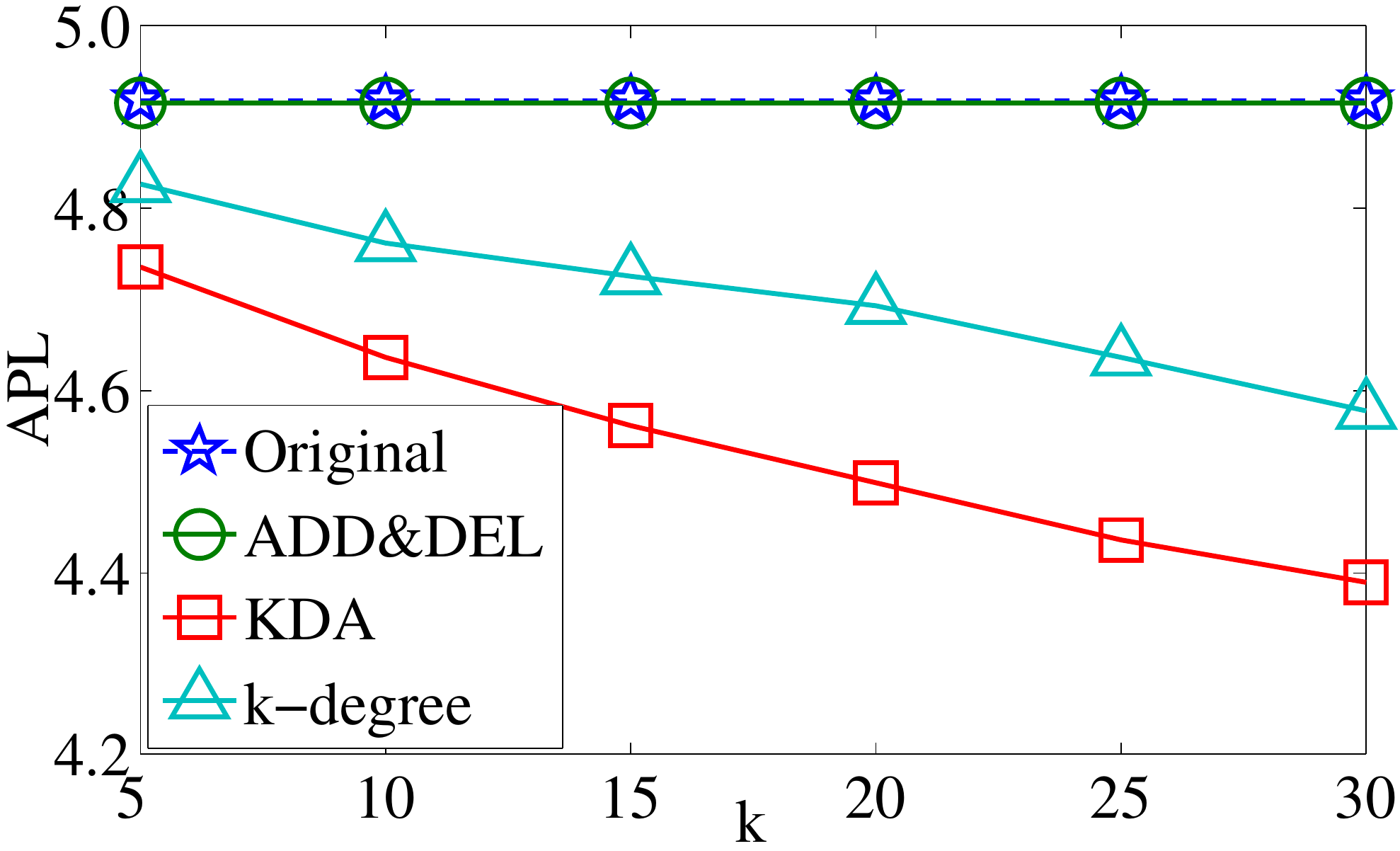}}}
\subfigure[BC]{\label{brightkite-deg-bc}
\raisebox{-0.2cm}{\includegraphics[width=0.3\textwidth, height=1.1in]{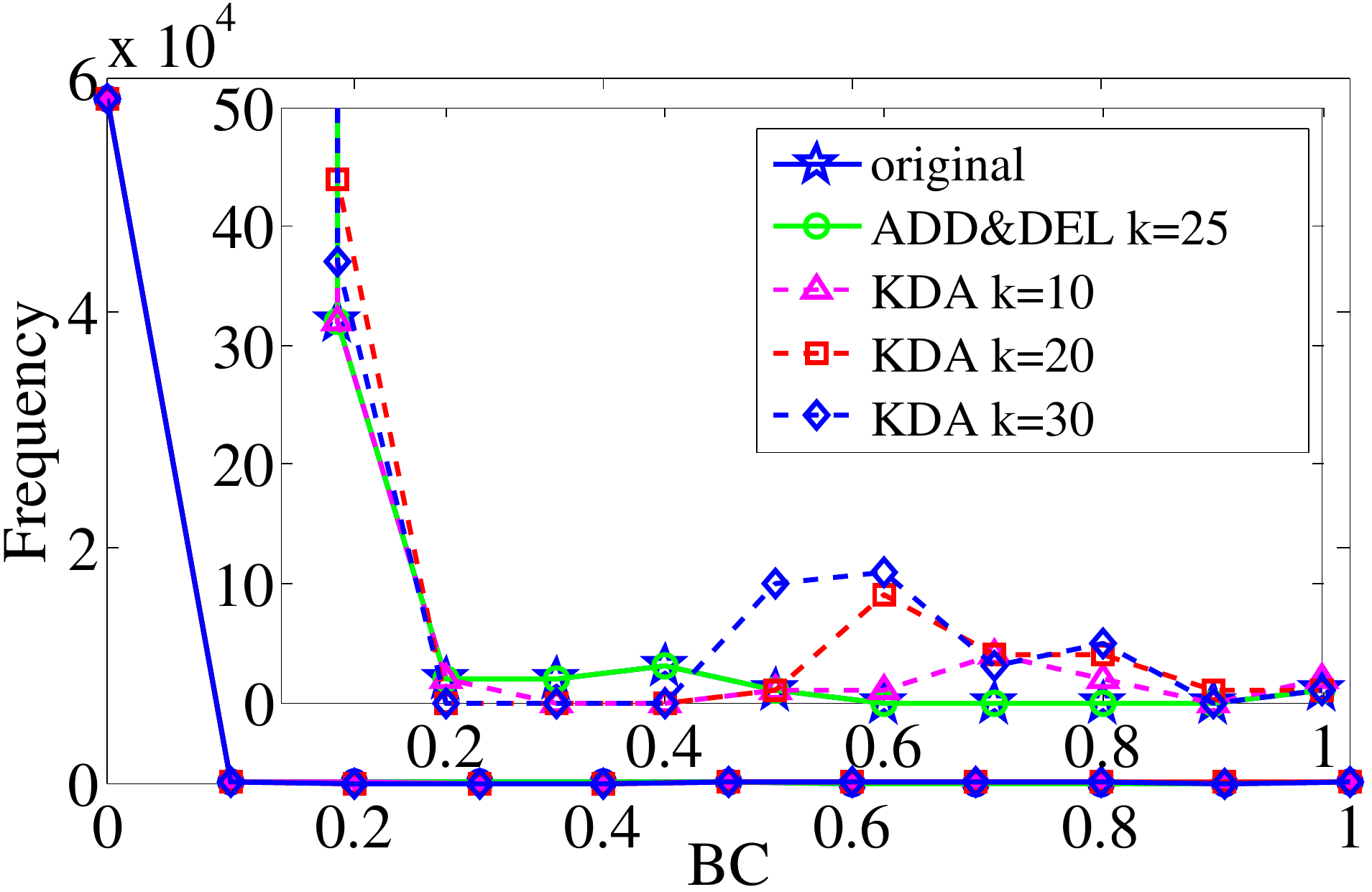}}}
\caption{k-degree anonymization on 25-NMF anonymized graph of Brightkite}\vskip -0.1in
\label{brightkite-deg}
\end{figure*}

Since there are no new triangles formed after the KDA algorithm adds new edges, the clustering coefficient decreases a little bit as $k$ increases as shown in Figures \ref{acm-deg-cc}, \ref{cora-deg-cc} and \ref{brightkite-deg-cc}. Our algorithm performs better than the classic $k$-degree anonymization on this measure.
Since new edges are added into the graph, the APL value decreases a little bit as $k$ increases as shown in Figures \ref{acm-deg-apl}, \ref{cora-deg-apl}, and \ref{brightkite-deg-apl}. As we consider the $k$-NMF anonymity, the classic $k$-degree anonymization performs a little better than our algorithm on the APL measure. But when the APL of the graph is large, our algorithm can perform better than the classic $k$-degree anonymization as shown in Figure \ref{acm-deg-apl}.
The results show that our algorithm performs well on preserving the utility while protecting the privacy by carefully exploring the graph property. The classic $k$-degree anonymization makes less effort on this except minimizing the number of edges added.
Figures \ref{acm-deg-bc}, \ref{cora-deg-bc} and \ref{brightkite-deg-bc} show the distributions of betweenness centrality of graphs anonymized by the KDA algorithm when we set $k_{deg}$ as 10, 20 and 30. The distributions of the anonymized graphs are very similar to the distributions of the original graphs especially for the ACM and Brightkite datasets. It shows that the KDA algorithm can preserve much of the utility of the graph anonymized by the $k$-NMF algorithms.

\vspace{-0.2cm}
\section{Conclusions}
In this paper, we have identified a new problem of $k$-anonymity on the number of mutual friends, which protects against the mutual friend attack in the social network publication. To solve this problem, we designed two heuristic algorithms which consider the utility of the graph. We also devised an algorithm to ensure the $k$-degree anonymity based on the $k$-NMF anonymity. The experimental results demonstrate that our approaches can ensure the $k$-NMF anonymity while preserve much of the utility in the original social networks.

\balance
\begin{spacing}{0.3} {\small
\bibliographystyle{IEEEtran}
\bibliography{MFA}
}
\end{spacing}

\end{document}